\documentclass[12pt]{iopart}

\usepackage[dvips]{graphicx}
\usepackage{pstcol,pst-node}

\eqnobysec

\newcommand{\beq}{\begin{equation}}
\newcommand{\eeq}{\end{equation}}
\newcommand{\ba}[1]{\begin{array}{#1}}
\newcommand{\ea}{\end{array}}
\newcommand{\bea}{\begin{eqnarray}}
\newcommand{\eea}{\end{eqnarray}}
\newcommand{\nn}{\nonumber \\}
\newcommand{\ben}{\begin{enumerate}}
\newcommand{\een}{\end{enumerate}}
\newcommand{\bit}{\begin{itemize}}
\newcommand{\eit}{\end{itemize}}
\newcommand{\bde}{\begin{description}}
\newcommand{\ede}{\end{description}}

\newcommand{\sz}{\scriptsize}

\newcommand{\sign}{\mbox{sign}}

\newcommand{\req}[1]{(\ref{#1})}

\definecolor{color1}{rgb}{0.0,0.5,1.0}
\definecolor{color2}{rgb}{0.0,1.0,0.0}
\definecolor{color3}{rgb}{0.0,0.0,1.0}
\definecolor{color4}{rgb}{0.0,0.5,0.0}

\begin{document}

\title{Analysis of the structure of complex networks at different resolution levels}

\author{A~Arenas$^{1,2,3}$\footnote{Author to whom any correspondence should be addressed},
A~Fern\'andez$^1$ and S~G\'omez$^1$}

\address{$^1$ Departament d'Enginyeria Inform\`{a}tica i Matem\`{a}tiques,
Universitat Rovira i Virgili,
Avinguda dels Pa\"{\i}sos Catalans 26,
43007 Tarragona, Spain}

\address{$^2$ Institute for Biocomputation and Physics of Complex Systems (BIFI),
Universidad de Zaragoza,
Corona de Arag\'on 42, Edificio Cervantes,
50009 Zaragoza, Spain}

\address{$^3$ Lawrence Berkeley National Laboratory, Berkeley, CA 94720, USA}

\eads{\mailto{alexandre.arenas@urv.cat},
\mailto{alberto.fernandez@urv.cat} and
\mailto{sergio.gomez@urv.cat}}

\begin{abstract}
Modular structure is ubiquitous in real-world complex networks, and its detection is important because it gives insights in the structure-functionality relationship. The standard approach is based on the optimization of a quality function, modularity, which is a relative quality measure for a partition of a network into modules. Recently some authors \cite{fortunato,kerstez} have pointed out that the optimization of modularity has a fundamental drawback: the existence of a resolution limit beyond which no modular structure can be detected even though these modules might have own entity. The reason is that several topological descriptions of the network coexist at different scales, which is, in general, a fingerprint of complex systems. Here we propose a method that allows for multiple resolution screening of the modular structure. The method has been validated using synthetic networks, discovering the predefined structures at all scales. Its application to two real social networks allows to find the exact splits reported in the literature, as well as the substructure beyond the actual split.
\end{abstract}

\pacno{89.75}
\submitto{\NJP}
\maketitle

\section{Introduction}
In 2002, Girvan and Newman \cite{firstnewman} highlighted the relevance of community structure in complex networks and proposed a method to detect it. This work opened a new scenario that has deserved a lot of attention in recent years \cite{newman_rev,jstat} specially because they identified structures which have meaning, they reveal information about roles of groups of nodes. This is the case, for example in the worldwide airports network \cite{rogerair}, the WWW \cite{flake}, biological networks \cite{holme,amaral,palla}, social networks \cite{firstnewman,danon} and the Internet \cite{eriksen,adamic}, among others. The information revealed by the community structure of real networks can be very valuable and make scientists aware of accuracy and reliability of the method used to detect this substructure.

The most important advance about community detection from the previous hit \cite{firstnewman} was given by the same authors \cite{newgirvan}, proposing a quality measure, {\em modularity} ($Q$), that allows to quantify the modular structure. Given a network partitioned into communities or modules, being $C_i$ the community to which node $i$ is assigned, the mathematical definition of modularity \cite{newanaly} is expressed in terms of the weighted adjacency matrix $w_{ij}$, that represents the value of the weight in the link between nodes $i$ and $j$ ($0$ if no link exists), and the strengths $w_i=\sum_j w_{ij}$ as
\beq
  Q=\frac{1}{2w}\sum_i\sum_j\left(w_{ij}-\frac{w_i w_j}{2w}\right)\delta(C_i,C_j)\,,
  \label{QW}
\eeq
where the Kronecker delta function $\delta(C_i,C_j)$ takes the values, 1 if node $i$ and $j$ are into the same module,  0 otherwise, and the total strength is $2w=\sum_i w_i$. For unweighted networks $w_i$ becomes the degree of node $i$, and $w$ the total number of links of the network.

The modularity of a given partition is the probability of having edges falling within modules in the network minus the expected probability in an equivalent (null case) network with the same number of nodes, and edges placed at random preserving the nodes' strength. The larger the modularity the best the partitioning is, because more deviates from the null case. Note that the optimization of the modularity cannot be performed by exhaustive search since the number of different partitions are equal to the Bell numbers \cite{bell}, which grow at least exponentially in the number of nodes $N$.
%Indeed, optimization of modularity is a NP-hard problem \cite{brandes}.
Heuristics for the optimization of modularity \cite{newfast,clauset,rogernat,duch,pujol,newspect} have become the only feasible (in computational time), and accurate method to detect modular structure up to now.

Recently, Fortunato and Barth{\'e}lemy \cite{fortunato} showed mathematically that the optimization of modularity has a resolution limit, raising important concerns about the reliability of the modules detected so far using this technique, or eventually using any other quality function. Using a definition of module extracted from the functional form of \req{QW} they subscribe the possible existence of undetectable submodules within the modules obtained optimizing \req{QW}. The same limitation has been observed for other quality functions different from modularity \cite{kerstez}.

Here we address the issue of community detection in two ways:  first, clarifying the conceptual interpretation of the resolution limit, not as a problem but as a feature of quality functions that can help us to understand in deep the structure of networks, and second and most important, we provide with a method that allows the full screening of the topological structure at any resolution level using the original definition of $Q$. Once presented the method, we will compare it with recent approaches intended to explore, also, the substructure of networks.

\section{Complex networks topology represented at different scales}

\subsection{Resolution limit and topological scales}
Rewriting \req{QW} in terms of contribution of modules instead of nodes we have
\beq
  Q=\sum_{s=1}^{m} \left(\frac{w_{ss}}{w}-\left(\frac{w_{s}}{2w}\right)^{2}\right),
  \label{QWS}
\eeq
where the sum is over the $m$ modules of the partition, $w_{ss}$ is the internal strength of module $s$ and $w_s$ the total strength of module $s$. For unweighted networks $w_{ss}$ reduces to the number of internal links and $w_s$ to the sum of degrees of the nodes in module $s$.

The solution we propose takes advantage of the dependence of the resolution limit on the total strength $2w$. Consider the case study consisting on two identical modules with a single link connecting them to the rest of the network and only one link connecting them to each other \cite{fortunato}, the resolution limit states that these modules will not be found, optimizing modularity, if their internal strengths are
\beq
  w_{ss}<\sqrt{w/2} - 1\,.
  \label{limit}
\eeq
In \cite{fortunato} the authors neglect the contribution $-1$ in the second side of inequality \req{limit}, which is acceptable for large values of the total strength.

Our proposal to solve this problem is to modify the total strength $2w$. Let us assume that we increase the strength of every node by a quantity say $r$, then \req{limit} will read
\beq
  w_{ss} <\frac{1}{2}\left(\sqrt{\left(2w+Nr\right)}-n_sr - 2\right),
  \label{limit_r}
\eeq
where $n_s$ stands for the number of nodes in module $s$ and $N$ for the number of nodes in the network. The result of this prescription resulting in \req{limit_r} is that by rescaling the topology by a factor $r$,  the example above can be separated optimizing modularity, because the growth of $\sqrt{r}$ is slower than $r$, i.e.\ at some scale controlled by $r$ both modules will be visible using optimal modularity.

The problem now is how to increase the strength of nodes without altering the topological characteristics of the original network. We solve this problem by rescaling the topology defining ${\bf W}_{r}$, from the original weighted adjacency matrix $\bf W$ of the graph with entries $w_{ij}$, as follows
\beq
  {\bf W}_{r}={\bf W} + r\bf{I}\,,
  \label{mat}
\eeq
where {\bf I} is the identity matrix. In terms of graphs, this new matrix represents the original network with self-loops of weight $r$ for every node. Note that the prescription in \req{mat} supposes a constant shift (translation) $r$ of the strength of each node.

The commonly analyzed structural characteristics of networks (strength distribution, weighted clustering coefficient, strength correlations of any order, etc.) remain the same in the new network because the translation of strengths does not affect the original links' weights $w_{ij}$ that are the building blocks of the topology. The shift only affects the property of each node individually and in the same way for all them. The spectra of the original graph is also shifted a quantity $r$ for each eigenvalue, preserving then any property that depends on differences between eigenvalues. The eigenvectors are exactly the same. Finally, the associated Laplacian matrix of the original matrix $L_{ij}=w_{i}\delta_{ij} - w_{ij}$, responsible for the behavior of linear dynamical processes on the network \cite{arenas}, is also unchanged.

The interesting property of the re-scaled topology ${\bf W}_{r}$ is that its characteristic scale in terms of modularity has changed. Then the topological structure revealed by optimizing modularity for ${\bf W}_{r}$ is that of large groups for small values of $r$, and smaller groups for large values of $r$, all them strictly embedded in the original topology. This fact allows for the screening  of the modular structure by analyzing the optimal modular structure of ${\bf W}_{r}$ for different values of $r$. Note that the rescaling of the topology is simply an elegant way to enhance the total strength of the network, without varying its topological properties, then the rescaling can be used, in principle, to analyze the structure of networks using any quality function at different resolution levels parametrized by $r$.

\subsection{Multiple resolution method}
The analysis of modules at different resolution levels that we propose, consists into optimize the modularity of the graph ${\bf W}_{r}$ for different values of $r$. Denoting $Q_r$ the modularity of the network at scale $r$, the equivalent expression to \req{QWS} reads
\beq
  Q_r=\sum_{s=1}^{m} \left( \frac{2w_{ss}+n_{s}r}{2w+Nr}-\left(\frac{w_{s}+n_{s}r}{2w+Nr}\right)^{2}\right).
  \label{QWSR}
\eeq

The topological scale determined by maximizing $Q$ at which the detection of modular structure has been attacked so far, corresponds to $r=0$. For positive values of $r$, we have access to the substructures underneath those at $r=0$, and for negative values of $r$ we have access to the superstructures. The topological scale corresponding to all nodes separated (forming their own communities) is found by maximizing $Q_{r_{\mbox{\tiny max}}}$, where $r_{\mbox{\sz max}}$ is the smallest positive value of $r$ that satisfies $w_{ij}<\frac{(w_i+r)(w_j+r)}{2w+Nr}$ for all $i \neq j$. And the topological scale corresponding to a unique module formed by the whole network is found by maximizing $Q_{r_{\mbox{\tiny min}}}$, where $r_{\mbox{\sz min}}$ has a lower bound defined by the asymptote
$r_{\mbox{\sz asymp}}=-\frac{2w}{N}$, for a detailed analysis see Appendix A. At the asymptote the total strength is zero, thus no meaningful scales can be found for values of $r$ below it. Note that the average strength can be written as $\frac{2w+Nr}{N} = r- r_{\mbox{\sz asymp}}$. To compare results at different resolution, we adopt the usual formulation in other areas of physics (optics, acoustics, etc.) where scales are prescribed as the logarithm of the ratio between the relevant parameter. Here, the difference between scales, is measured as the logarithm of the ratio between strengths
$\log(\frac{2w+Nr}{2w+Nr'}) \equiv \log(\frac{r-r_{\mbox{\tiny asymp}}}{r'-r_{\mbox{\tiny asymp}}})$.

In this new description, we have that a module is defined at each scale of description $r$, as the result of the maximization of $Q_r$. Moreover, modules that exist at a certain level of description may  disappear from our observation when changing the scale $r$ while others arise. Note that nothing implies that the substructures to which we will have access at different resolution levels are necessarily hierarchical, indeed in general they will not be hierarchical. Although, in principle, all resolution scales provide some information about the topology, and are important, the detection of partitions that are more persistent than the rest when changing the resolution $r$ is indicative of a tougher modular structure.

\section{Results}\label{sec:results}
We show the results of our method investigating the modular structure at multiple resolution levels (different scales), for examples of synthetic and real complex networks. A first approach on synthetic networks is illustrative for validation of the procedure when different coexistent  topological scales are imposed by construction. We have also analyzed the modular structure of real networks. In general, in real cases, the results are more difficult to assess because nothing from the topology indicates the existence {\em a priori} of more relevant structure in the network, and only the corroboration {\em a posteriori} of the structure found with known facts about the (social, biological, etc.) meaning of it can give reliability to any method. In the experiments, we have studied between 100 and 500 values of $r$ inside the interval $(r_{\mathrm{asymp}}, r_{\mathrm{max}}]$ for synthetic networks, and 1000 values of $r$ for real networks. All the experiments have been cross checked using two modularity optimization heuristics: extremal optimization \cite{duch}, and a new proposal for the optimization of modularity based on tabu search (see Appendix B for details), repeating each one 20 times and keeping the partition obtained at the optimal value of $Q_r$.

\begin{figure}[t]%figure1
  \begin{indented}
  \item[]
  \begin{tabular}[t]{ccc}
    \multicolumn{1}{l}{(a)} \\ \\
    \mbox{\includegraphics*[width=.34\textwidth]{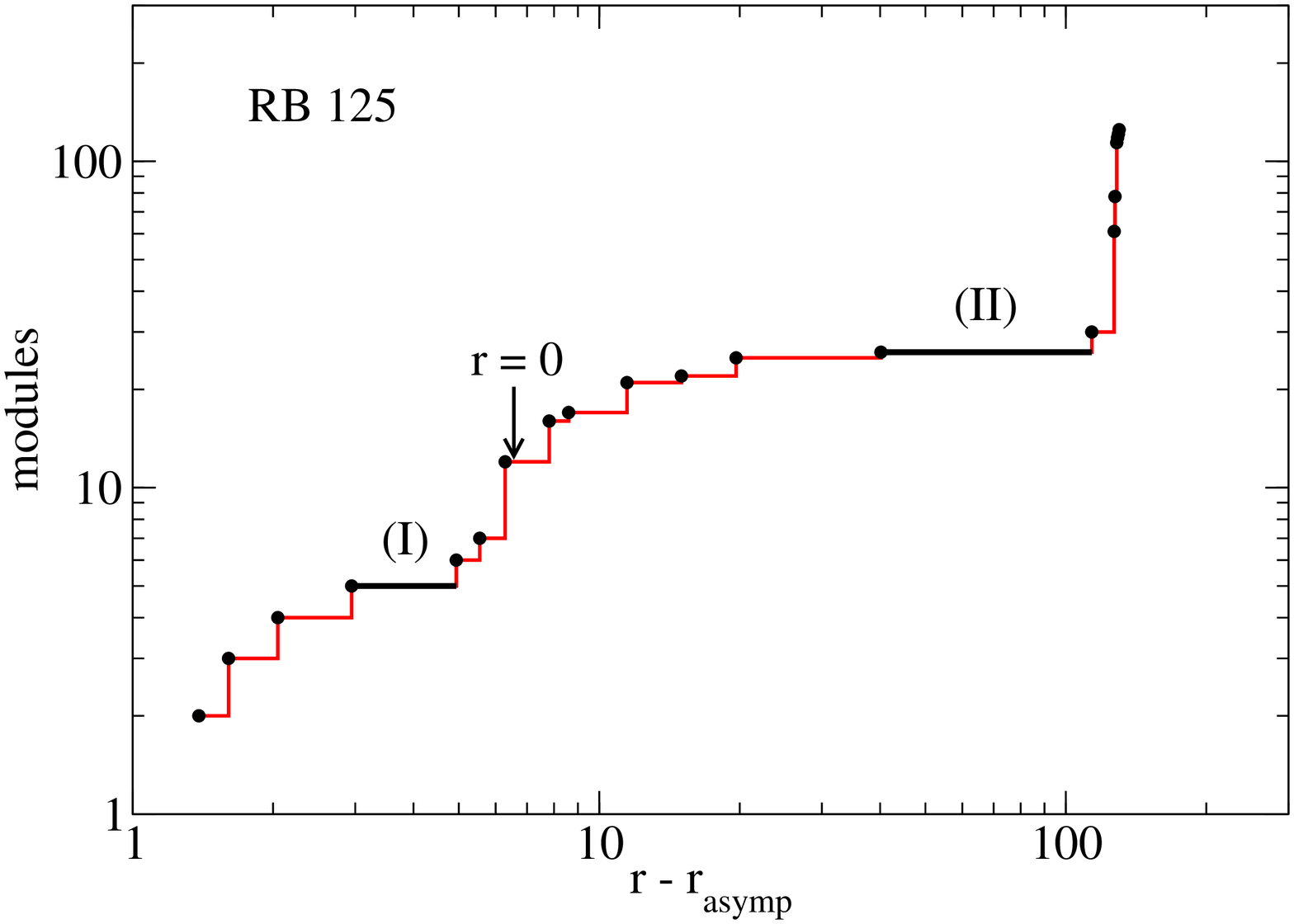}}
    &
    \begin{tabular}[b]{c}
    \mbox{\includegraphics*[width=.19\textwidth]{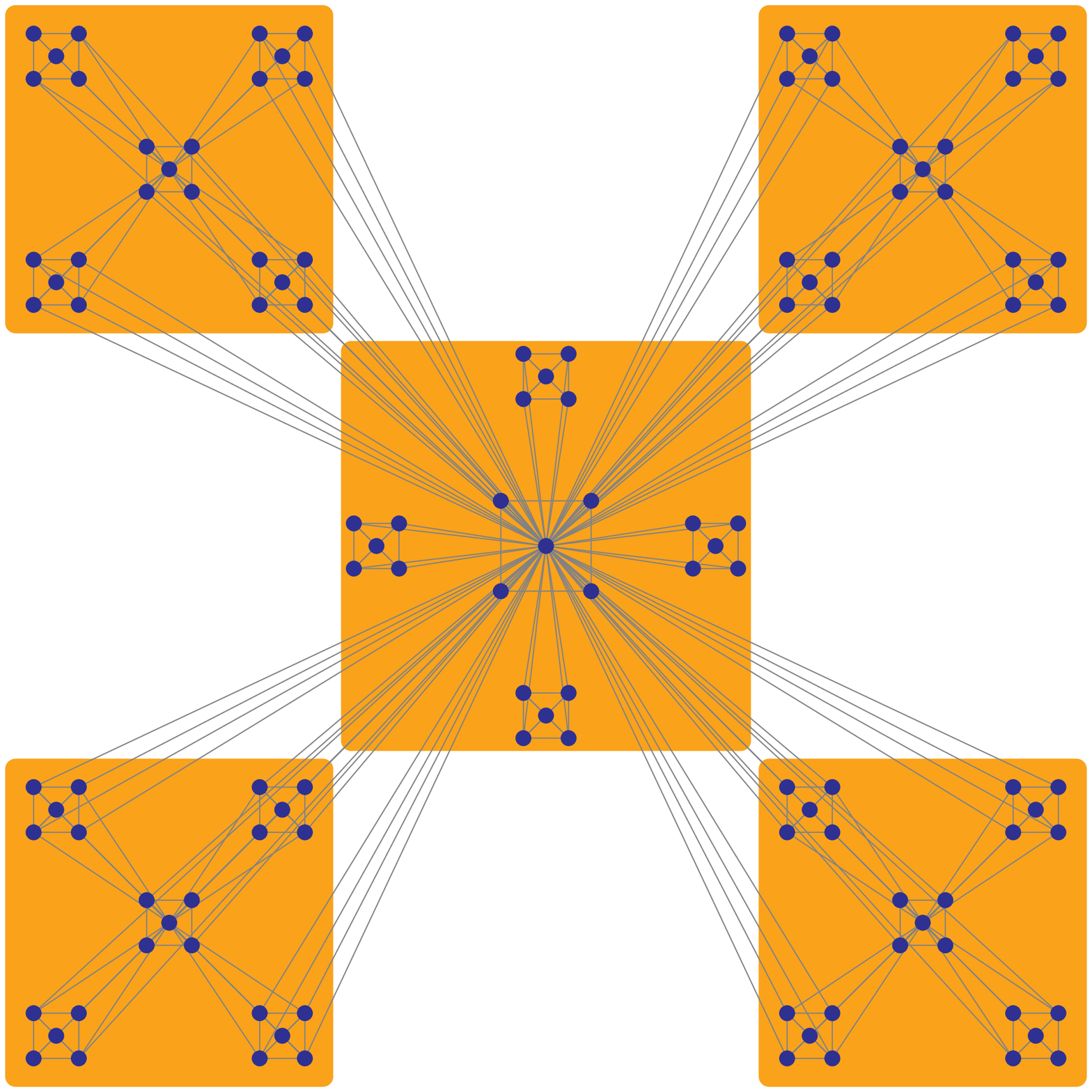}}
    \\ \mbox{\rule{0pt}{12pt}{\rm (I)}}
    \end{tabular}
    &
    \begin{tabular}[b]{c}
    \mbox{\includegraphics*[width=.19\textwidth]{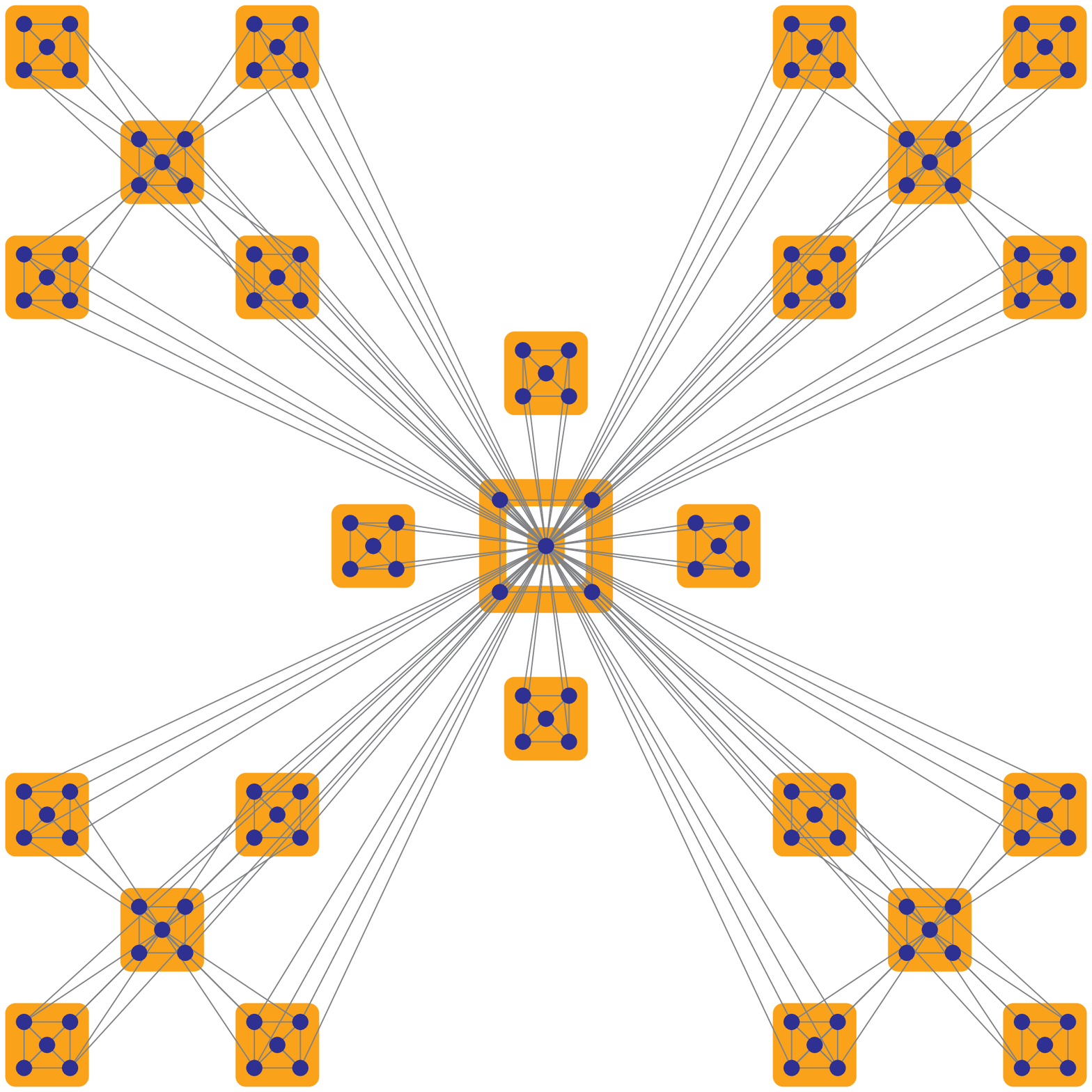}}
    \\ \mbox{\rule{0pt}{12pt}{\rm (II)}}
    \end{tabular}
    \\ \\
    \multicolumn{1}{l}{(b)} \\ \\
    \mbox{\includegraphics*[width=.34\textwidth]{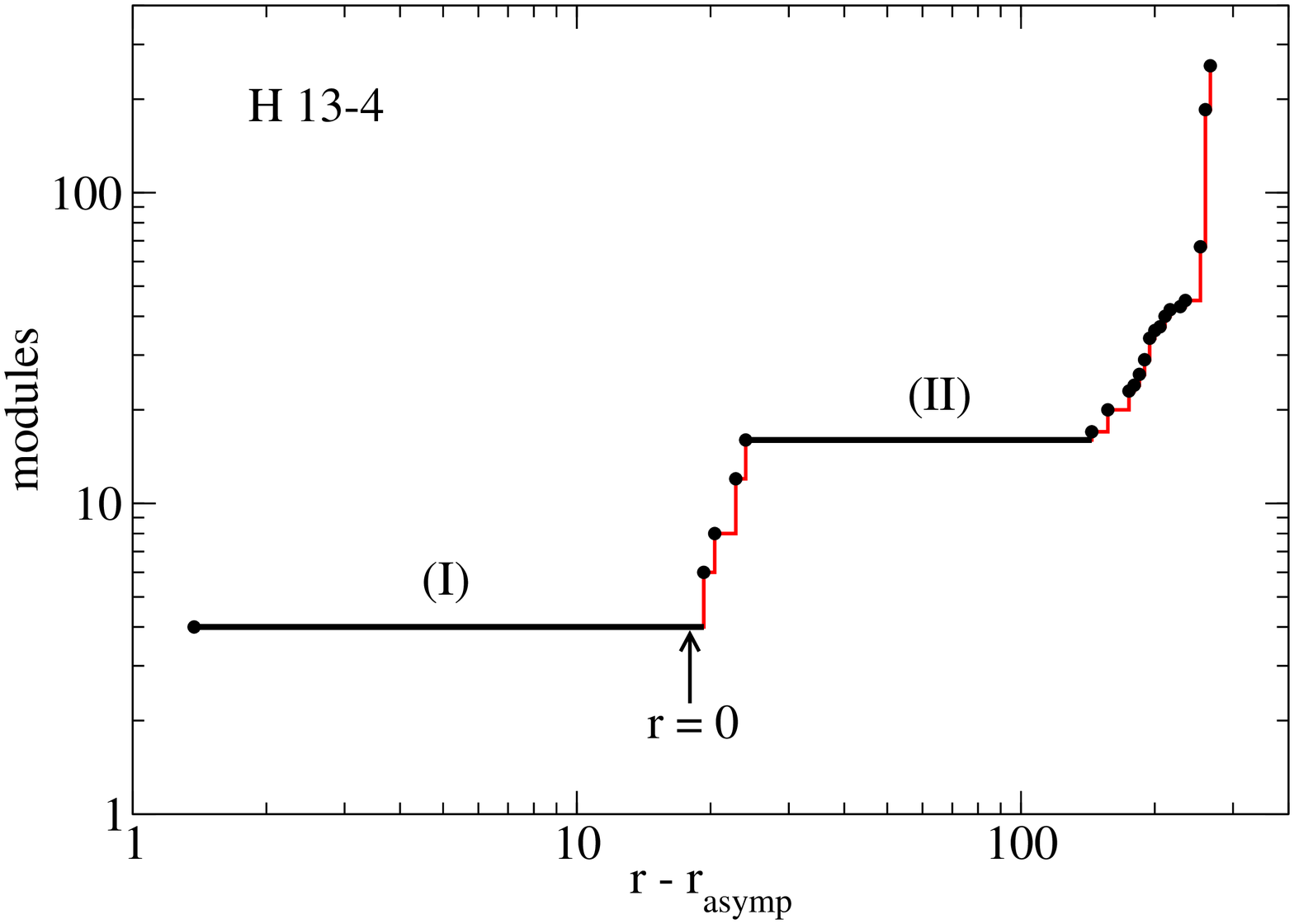}}
    &
    \begin{tabular}[b]{c}
    \mbox{\includegraphics*[width=.19\textwidth]{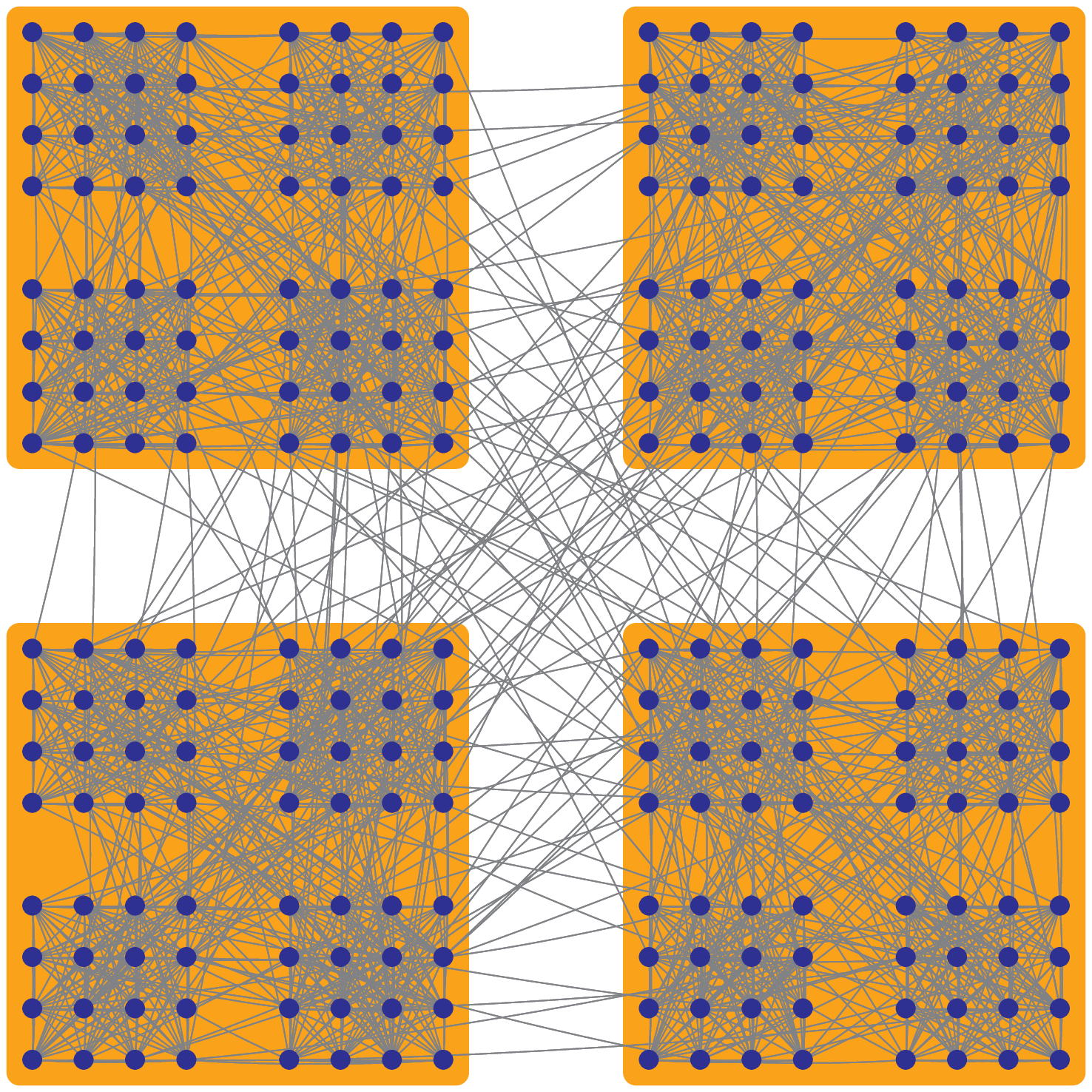}}
    \\ \mbox{\rule{0pt}{12pt}{\rm (I)}}
    \end{tabular}
    &
    \begin{tabular}[b]{c}
    \mbox{\includegraphics*[width=.19\textwidth]{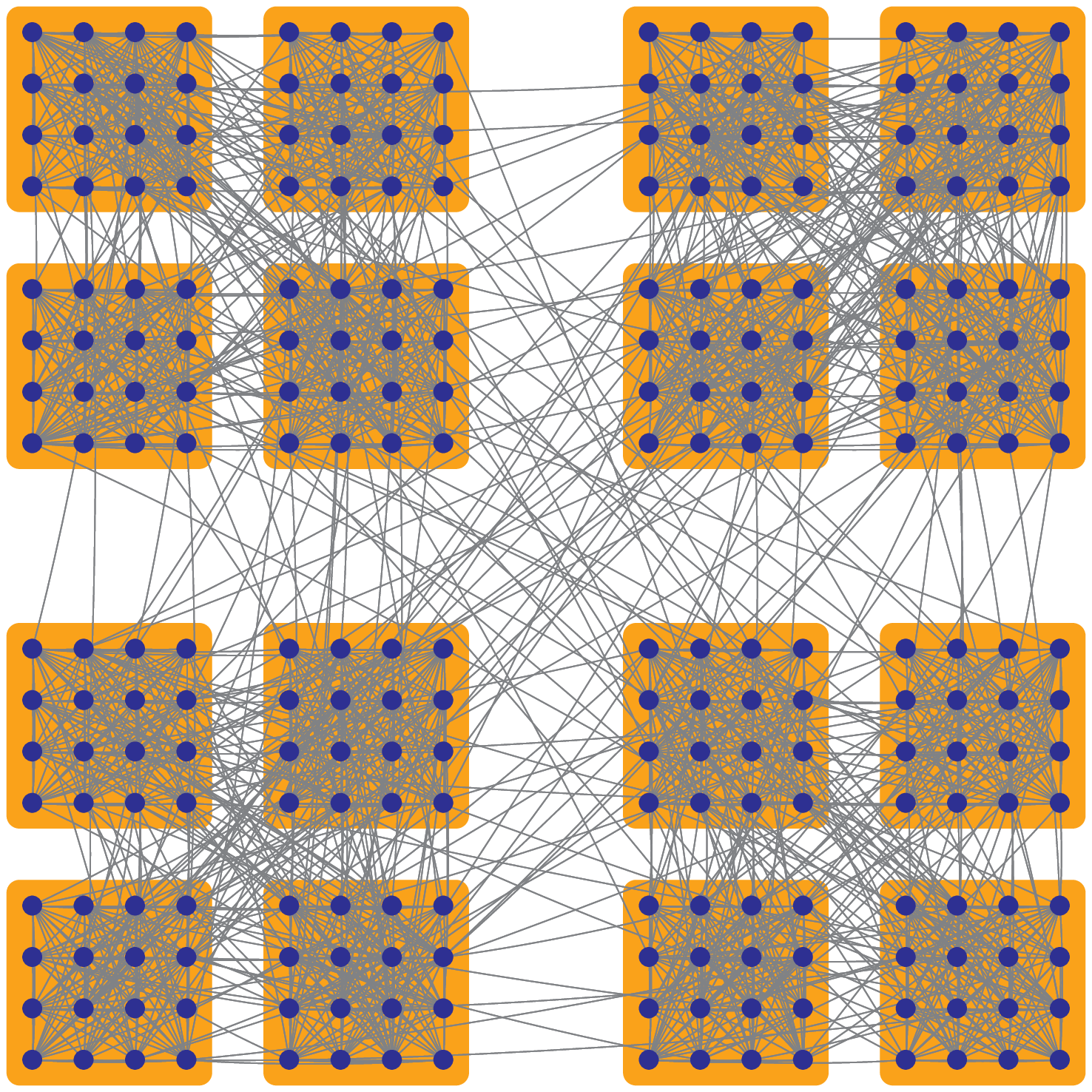}}
    \\ \mbox{\rule{0pt}{12pt}{\rm (II)}}
    \end{tabular}
    \\ \\
    \multicolumn{1}{l}{(c)} \\ \\
    \mbox{\includegraphics*[width=.34\textwidth]{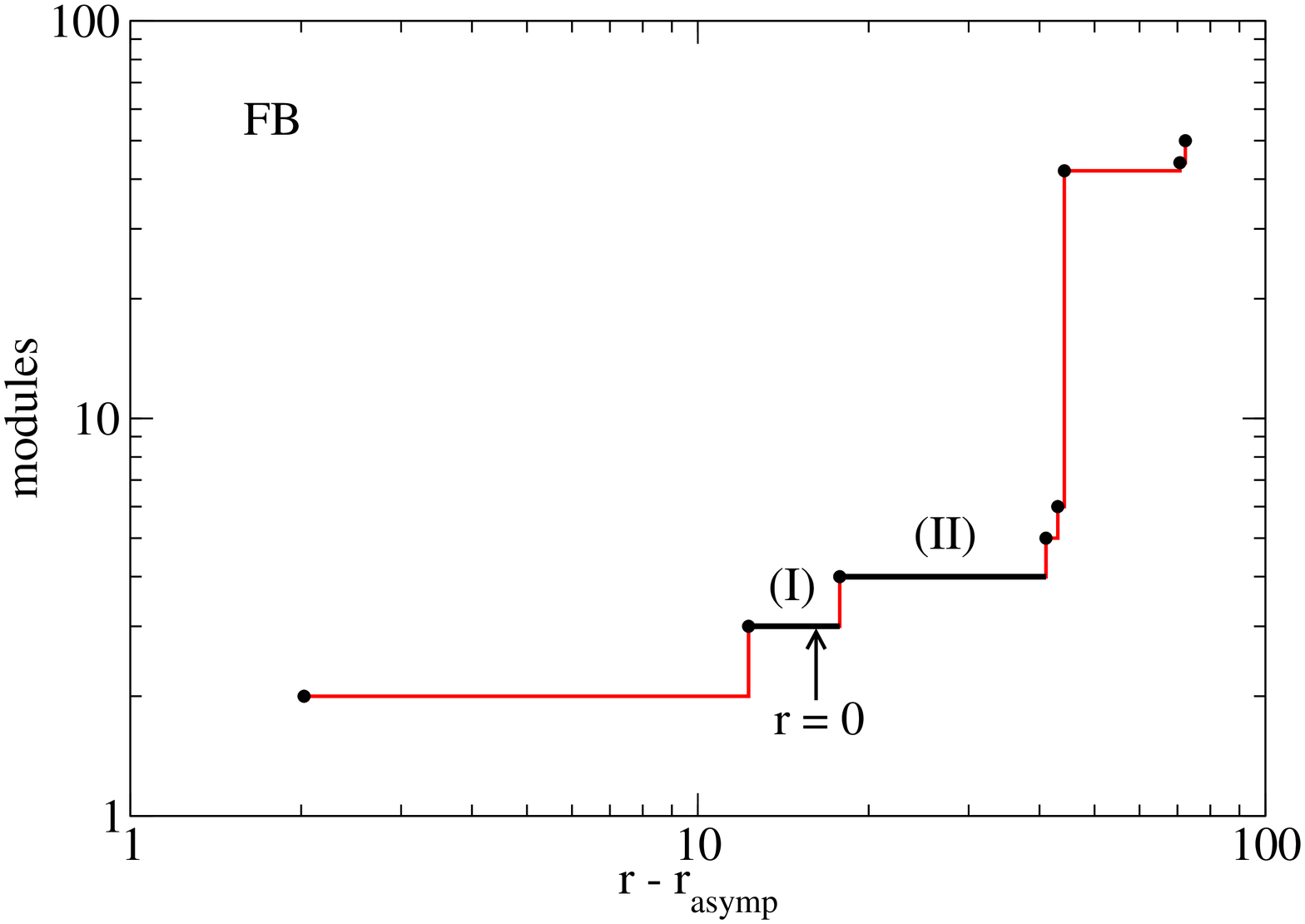}}
    &
    \begin{tabular}[b]{c}
    \mbox{\includegraphics*[width=.19\textwidth]{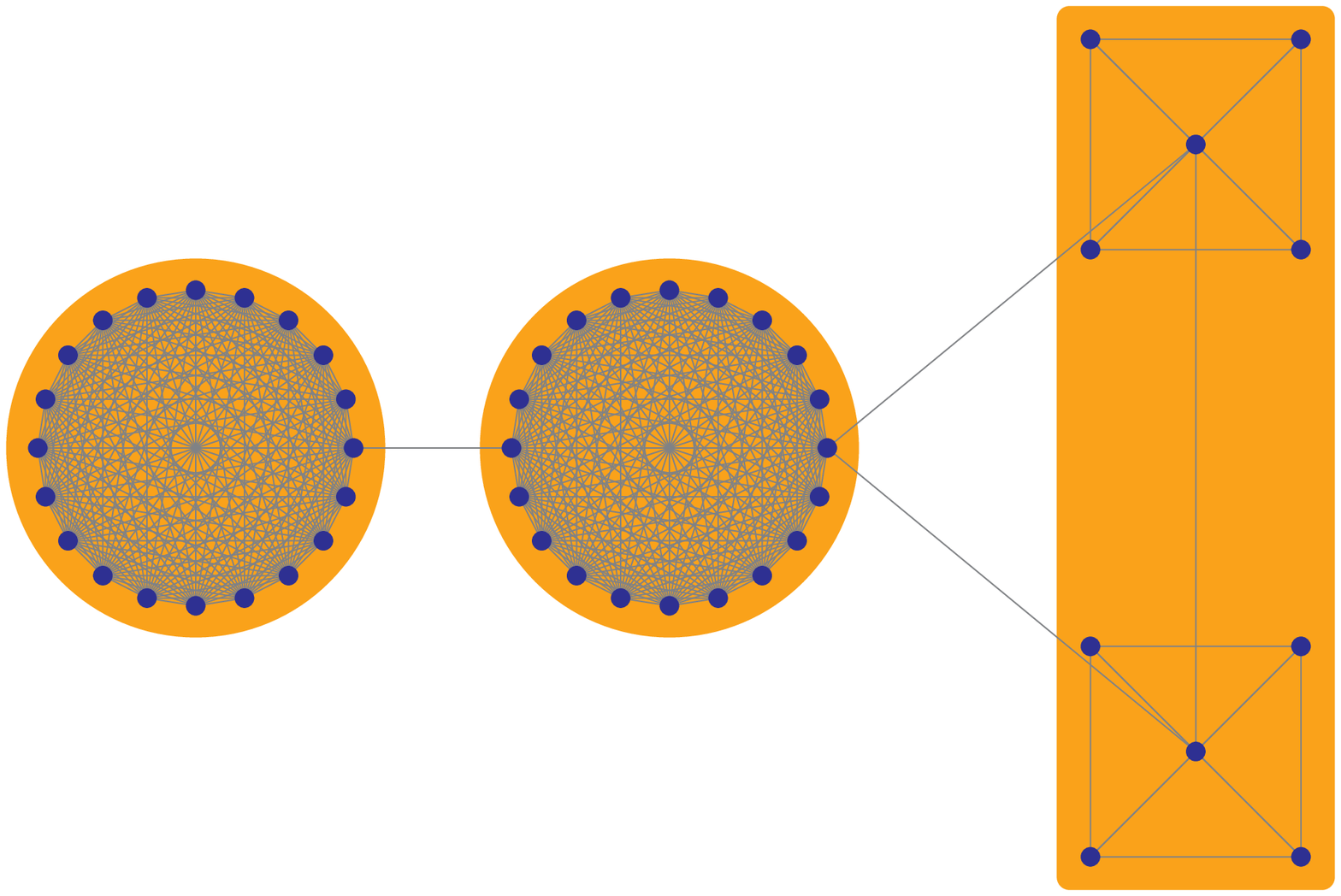}}
    \\ \mbox{\rule{0pt}{12pt}{\rm (I)}}
    \end{tabular}
    &
    \begin{tabular}[b]{c}
    \mbox{\includegraphics*[width=.19\textwidth]{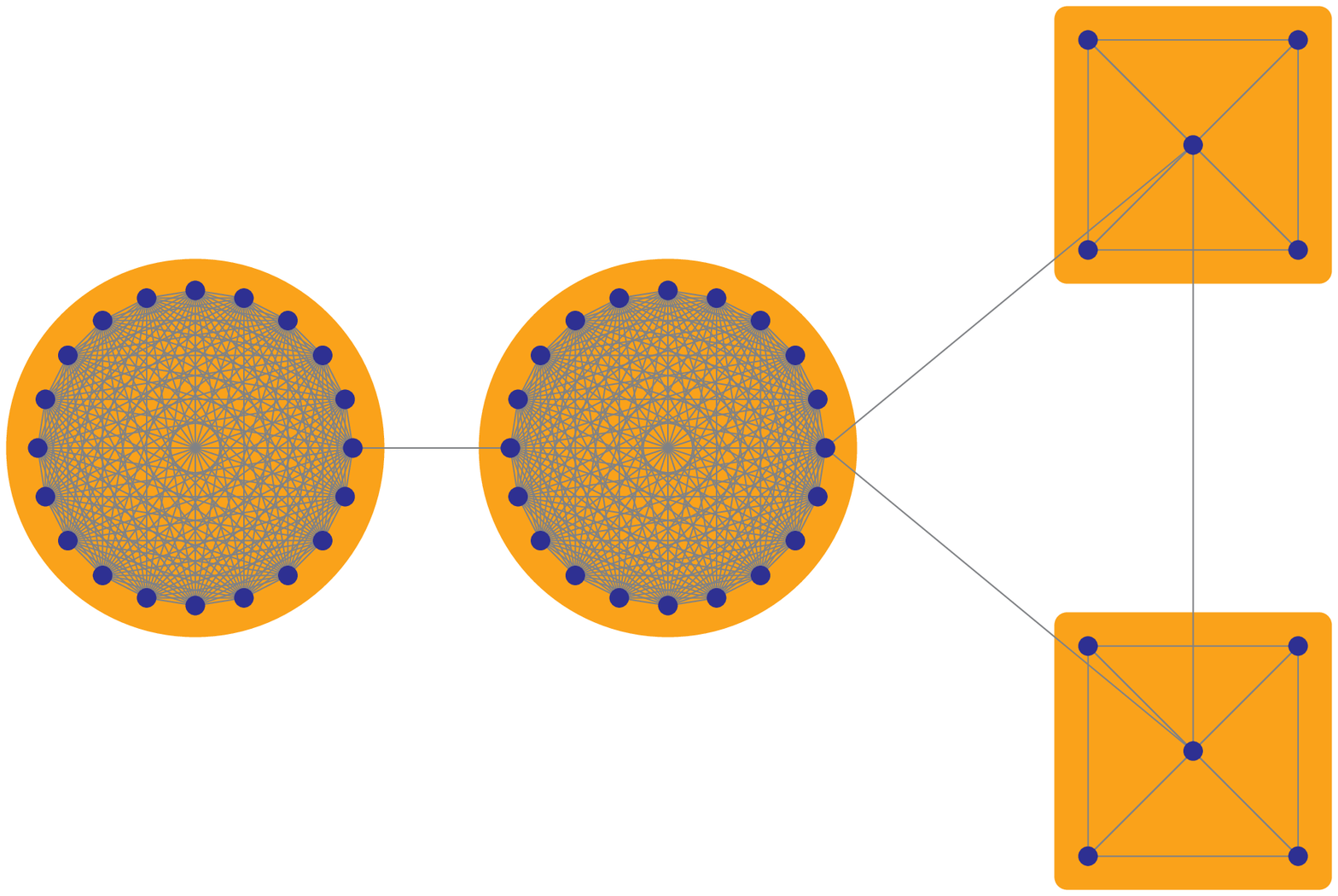}}
    \\ \mbox{\rule{0pt}{12pt}{\rm (II)}}
    \end{tabular}
  \end{tabular}
  \end{indented}
  \caption{Multiple resolution of modular structure in synthetic networks. Left: number of modules obtained at the optimal partition for $Q_r$, where each point corresponds to a different partition and the arrows indicate the optimal partitions at $r=0$. Right: networks analyzed, highlighting the partitions at two representative scales indicated by (I) and (II). (a) RB~125 corresponds to the hierarchical scale-free network proposed in \cite{rb}. The regions corresponding to 5, 25 and 26 modules are the most representative (stable) in terms of resolution. (b) H~13-4 corresponds to a homogeneous in degree network with two predefined hierarchical levels. Both levels are revealed by the method at different scales. (c) FB corresponds to the network proposed in \cite{fortunato} to demonstrate the resolution limit of modularity (at $r=0$). This limit is overcome at scale (II) providing with the partition expected.}
  \label{fig1}
\end{figure}

In figure~\ref{fig1} we have screened the whole range of topological scales for three synthetic networks, representing the number of modules obtained at the optimal partition for $Q_r$, and the network analyzed highlighting the partition at two representative scales indicated by (I) and (II). Although the networks studied may have more than two relevant scales, we have just drawn two of them chosen among the most representative ones. First we have computed the modular structure in a hierarchical scale-free network with 125 nodes, RB~125, proposed by Ravasz and Barabasi \cite{rb}. In figure~\ref{fig1}a we plot the modular structure found, which shows three different scales that deserve discussion. We observe clearly persistent structures in 5 and 25 communities respectively, that account for the subdivisions more significant in the process, showing two hierarchical levels for the structure. Additionally, the most stable partition in terms of resolution does not correspond to any of the previous ones, but it corresponds to the partition in 26 modules (the same as the one in 25 modules, but isolating the main hub). The partition in 5 modules and the partition in 26 modules are highlighted on the original network. This result is in perfect correspondence with the synchronization patterns produced on this network using coupled oscillators \cite{arenas}.

Another network example used is the H~13-4 network \cite{arenas}, which corresponds to a homogeneous in degree network with two predefined hierarchical levels, being 256 the number of nodes, 13 the number of links of each node with the most internal community (formed by 16 nodes), 4 the number of links with the most external community (four groups of 64 nodes), and 1 more link with any other node at random in the network. In figure~\ref{fig1}b we represent the network and its corresponding modular structure at different scales. Both hierarchical levels are revealed by the method as they correspond to the original construction of the network: the first hierarchical level consisting in 4 groups of 64 nodes, and the second level consisting in 16 groups of 16 nodes.

Finally, we have used the FB network proposed by Fortunato and Barth{\'e}lemy \cite{fortunato} to demonstrate the resolution limit of modularity (at $r=0$). It consists in two cliques of 20 nodes linked with two small cliques of 5 nodes. At $r=0$ the best partition cannot separate the two small cliques.  In figure~\ref{fig1}c we observe that the partition searched by the authors, formed by the four cliques isolated in their own communities, is obtained by increasing the resolution $r$, showing that the resolution limit of modularity is overcome by the method in region (II).

\begin{figure}[t]%figure2
  \begin{indented}
  \item[]
  \begin{tabular}[t]{cc}
    \multicolumn{1}{l}{(a)}
    &
    \multicolumn{1}{l}{(b)}
    \\ \\
    \mbox{\includegraphics*[width=.37\textwidth]{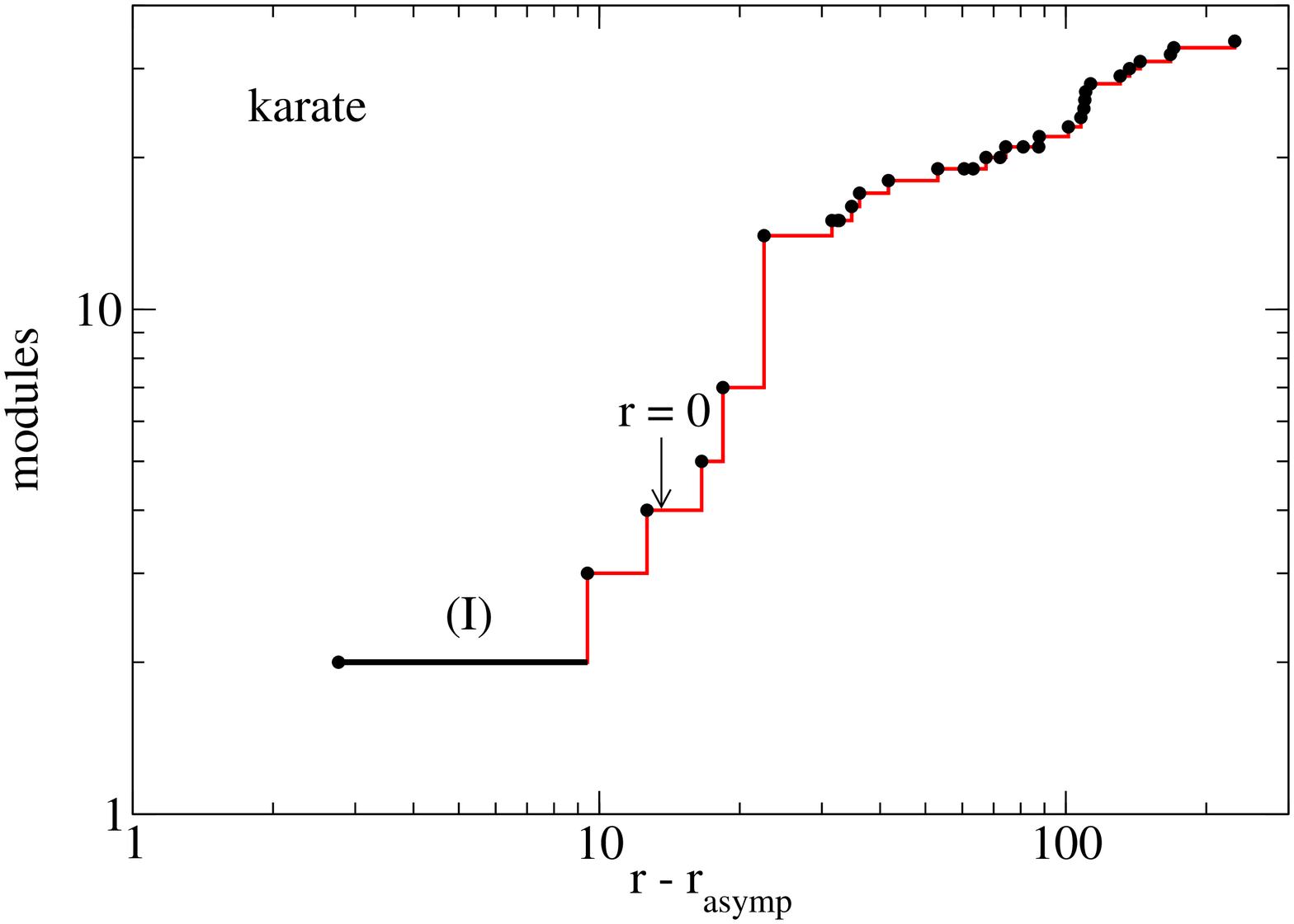}}
    &
    \mbox{\includegraphics*[width=.37\textwidth]{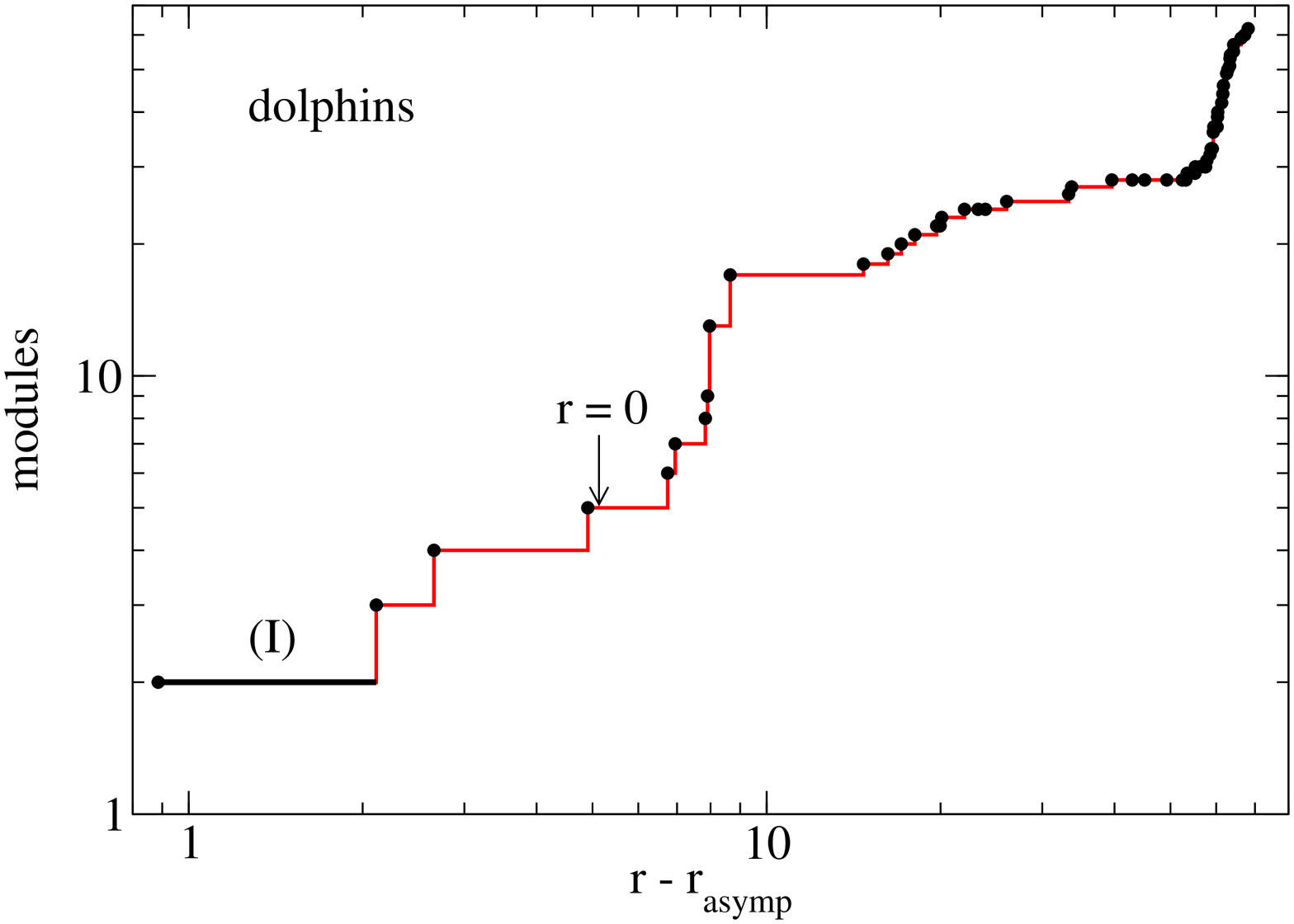}}
    \\ \\
    \begin{tabular}[b]{c}
    \mbox{\includegraphics*[width=.37\textwidth]{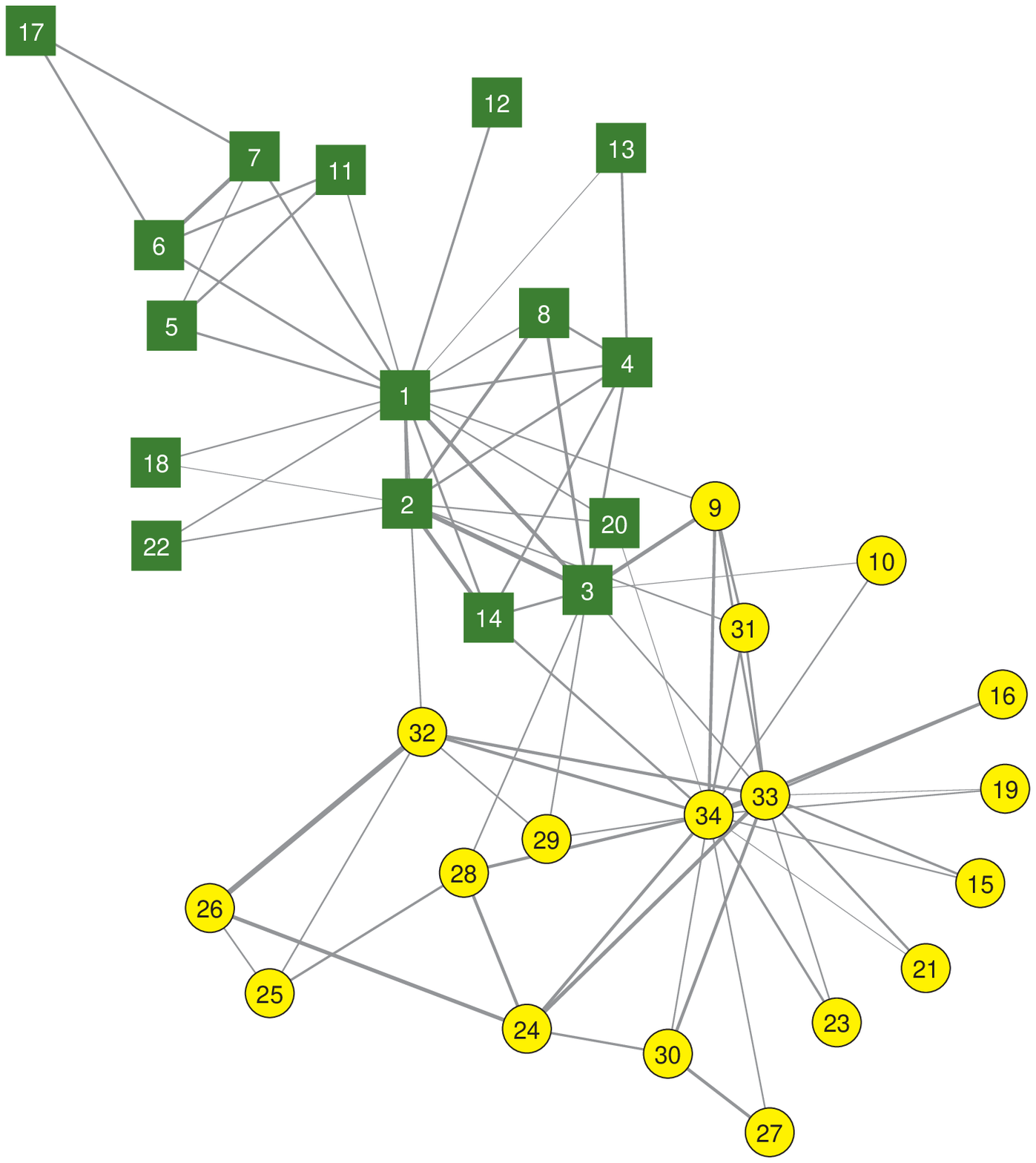}}
    \\ \mbox{\rule{0pt}{60pt}}
    \end{tabular}
    &
    \mbox{\includegraphics*[width=.37\textwidth]{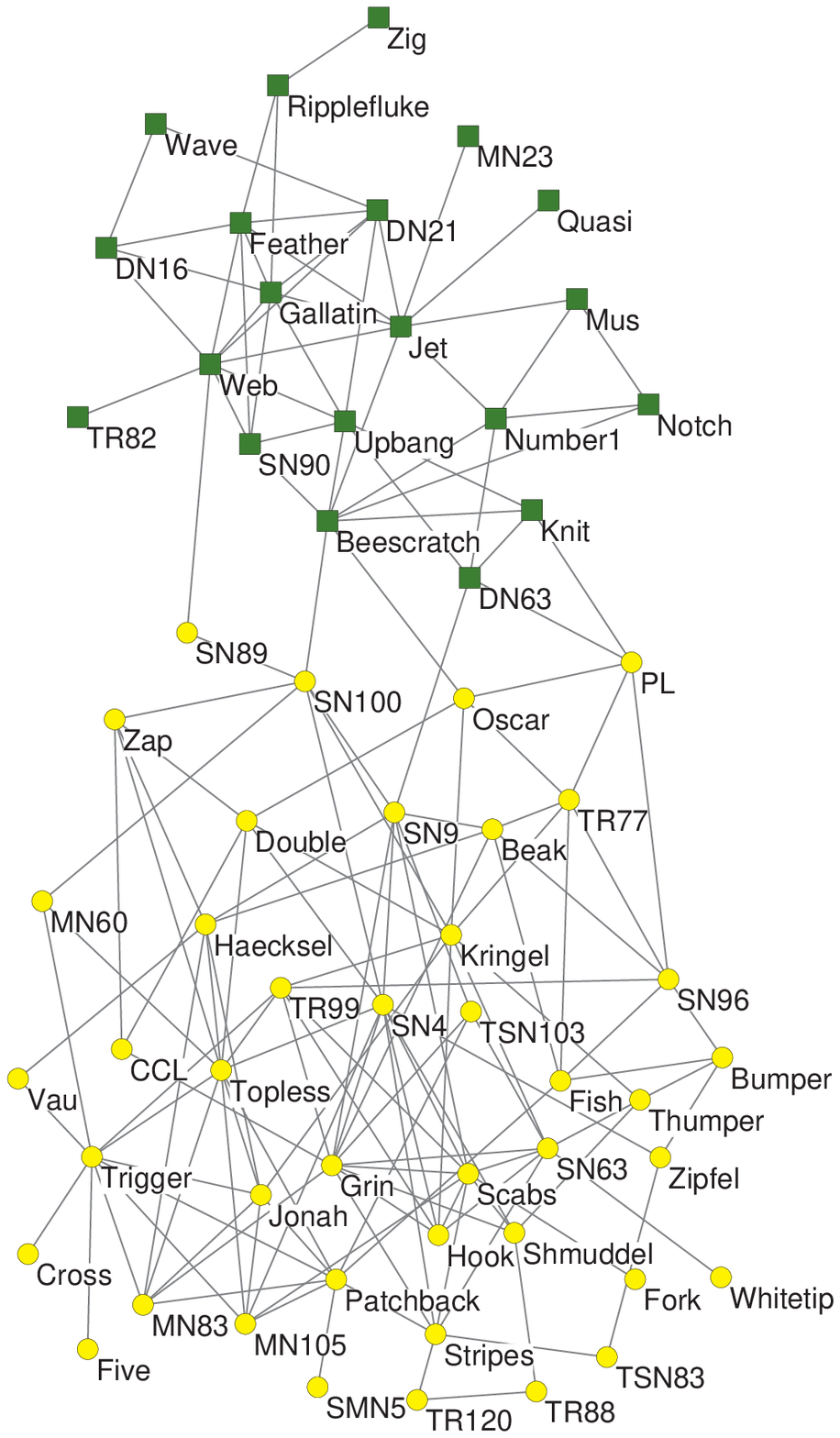}}
  \end{tabular}
  \end{indented}
  \caption{Multiple resolution of modular structure in real networks. Top: number of modules in the optimal partitions at different scales; the arrows indicate the best partitions obtained at $r=0$, which do not correspond to the real partitions. Bottom: representation of the networks and the partitions in the plateaus marked as (I), which correspond both to the most stable scales of description and to the known splittings occurred in the real networks. (a) Zachary's karate club network \cite{zachary}. (b) Dolphins social network by Lusseau et al.\ \cite{lusseau}.}
  \label{fig2}
\end{figure}

We have also studied a couple of social networks for which explicit knowledge about its modular structure is available, see figure~\ref{fig2}. These particular networks, formed by social acquaintances between individuals, have the main characteristic that after a period of study decomposed in perfectly identifiable parts. The challenge is to find the modular structure of these parts without previous knowledge about the real partition. The optimization of modularity at $r=0$ fails to provide this information, and no other method has been able to find the real partitioned structure. However, the most representative scales in terms of resolution optimizing $Q_r$ obtained by applying our method correspond exactly to the real splittings.

First, we have investigated the classical social network of the Zachary's karate club \cite{zachary}, accounting for the study over two years of the friendships between 34 members of a karate club at a US university in 1970. The network in question was divided, at the end of the study period, in two groups after a dispute between the club's administrator and the club's instructor, which ultimately resulted in the instructor leaving and starting a new club, taking about half of the original club's members with him. The analysis of this data has been a paradigmatic benchmark to test the accuracy of community detection algorithms. Zachary constructed a weighted network using different social measures, see figure~\ref{fig2}a, although many times in the physics literature the network has been considered unweighted for simplicity or tradition, missing important information in the process.

The goal of any community detection algorithm trying to identify modules on this network should be to find the actual split occurred, assigning perfectly the nodes to the two known resulting clubs. The first approach to this goal was given by Girvan and Newman in \cite{firstnewman}, where they used a divisive method that produced a hierarchical tree representing the whole modular structure. They found that the first network splitting obtained by the method assigned correctly all nodes except node number~3. However, no measure about the quality of the partition was introduced at that time, and then all levels of the hierarchical tree were equivalent, with no way to have a preference for any partition. In \cite{newgirvan}, the same authors introduced the modularity measure $Q$ and reported that the best structure in the hierarchy, in terms of the value of $Q$, was a partition in four groups and not two as expected. From this point on, many authors have analyzed this network and have provided the best values of $Q$ obtained. Today it is well accepted that the best partition in terms of modularity of the Zachary's unweighted network is achieved for four groups with a value of $Q=0.419$. We have applied our method to screen the modular structure of the original weighted network at all resolution scales of $r$. The results in figure~\ref{fig2}a show that the most stable level of resolution is precisely the partition resulting in the two groups representing the two clubs, with no mismatch of any individual.

The second network analyzed is the dolphins social network of Lusseau et al.\ \cite{lusseau}. This network was constructed from observations of a community of 62 bottle nose dolphins over a period of seven years from 1994 to 2001. The nodes in the network represent the dolphins, and the ties between nodes represent the associations between dolphin pairs occurring more often than expected by chance. There is evidence \cite{lunew} that a temporary disappearance of the dolphin denoted SN100, led to the fission of the dolphins community in the two identifiable parts shown in figure~\ref{fig2}b. The optimization of modularity at $r=0$ does not produce the expected split, but a partition in five communities with $Q= 0.518$. Other approaches as the one exposed in \cite{newspectpre} neither success to find the real division. Our method allows to reveal all the modular structure in the whole range of resolution, indicating that the most stable solution in terms of resolution of optimal $Q_r$ corresponds exactly to the two partitions observed in this animal social network.

\section{Discussion}
The variation of topological scales so far is mediated by the variation of the parameter $r$. The meaning of this parameter is that of a {\em resistance} to become part of a community, in the scope of modularity. For positive values of $r$, we have access to the substructures below those at $r = 0$ (corresponding to the original scale at which modularity was defined by Newman), and for negative values of $r$ we have access to the superstructures. The screening of different scales of descriptions should be useful to get deeper in the understanding of complex networks. Here we present a discussion about the role of the different topological scales beyond its statical definition, revealing their implications in dynamical process on top of networks. After that we will also compare our method with another possible approach to the mesoscale, and finally we give a perspective about the significance of the mesoscale in contrast with the commonly accepted one-scale of description.

\subsection{The contact with physics}
The results show that there exist several intermediate scales of description of the complex networks, the topological mesoscale. These scales are revealed by intervals of values of the resistance $r$, for which the optimal partition does not change, see figure~\ref{fig1}. The obvious question at this point is: what are these scales representative for? The answer of this question is not trivial, and is intrinsically related to the functioning of the complex network as a substrate for different dynamical processes, communication and friendship in social networks, cognitive task in neural networks, or different levels of aggregation of computers in the Internet, for example. Our guess is that a simple dynamical process on top of a complex networks, should somehow reveal the topological mesoscale also in terms of temporal patterns. To check this hypothesis, we have implemented a synchronization dynamics on top of different topologies following \cite{arenas,arenas1}. The dynamics corresponds to the non-linear interaction between oscillators connected following the links of the complex networks. Analyzing the temporal meta-stable patterns emerging in the evolution towards complete synchronization, we corroborate our initial guess.

The temporal mesoscale of the dynamics of synchronization (of phase oscillators) near the synchronization attractor are governed by the solutions of the linear dynamics:
\beq
  \frac{d\theta_i}{dt}=-k\sum_{j}
  L_{ij} \theta_j \,, \ \ i=1,\ldots,N\,,
  \label{linearmodel}
\eeq
where $k$ is a constant, $ \theta_j$ are the phases of the nodes and $L_{ij}$ the Laplacian matrix of the network.

To identify patterns of synchronization in time, we use \cite{arenas} a discretization of the matrix $\rho_{ij}=\langle cos(\theta_i-\theta_j)\rangle$ where $\langle\cdots\rangle$ stands for the average over different realizations of the initial conditions. In all cases presented here we have averaged $10^5$ realizations, and used a discretization threshold of 0.999. We observe that the intermediate scales that are revealed by the synchronization process are in agreement with those found by the topological method proposed here. The method allows not only to identify the number of communities at different scales but also to determine which nodes form these communities.

\begin{figure}[t]%figure3
  \begin{indented}
  \item[] (a)
  \item[]
    \mbox{}\hfill
    %\mbox{\includegraphics*[width=.6\textwidth]{fig3a.eps}}
    \mbox{

      \begin{pspicture}(-4.50,-2.00)( 4.50, 3.00)
      \tiny

      %\psframe[linecolor=red](-4.50,-2.00)( 4.50, 3.00)

      \pnode( 0.00,-0.59){A0}
      \pnode(-0.70,-1.79){A1}
      \pnode( 0.70,-1.79){A2}
      \pnode(-0.60,-1.34){A3}
      \pnode( 0.60,-1.34){A4}

      \pnode(-2.61, 1.81){B0}
      \pnode(-4.00, 0.00){B1}
      \pnode(-2.00, 0.00){B2}
      \pnode(-3.10, 0.90){B3}
      \pnode(-1.55, 0.90){B4}

      \pnode( 2.61, 1.81){C0}
      \pnode( 2.00, 0.00){C1}
      \pnode( 4.00, 0.00){C2}
      \pnode( 1.55, 0.90){C3}
      \pnode( 3.10, 0.90){C4}

      \pnode(-1.50, 2.76){D0}
      \pnode(-2.14, 1.85){D1}
      \pnode(-1.08, 1.85){D2}
      \pnode(-1.86, 2.14){D3}
      \pnode(-0.92, 2.14){D4}

      \pnode( 1.50, 2.76){E0}
      \pnode( 1.08, 1.85){E1}
      \pnode( 2.14, 1.85){E2}
      \pnode( 0.92, 2.14){E3}
      \pnode( 1.86, 2.14){E4}

      \ncline[linewidth=0.5pt,linecolor=gray,linecolor=gray]{A0}{B1}
      \ncline[linewidth=0.5pt,linecolor=gray]{A0}{B2}
      \ncline[linewidth=0.5pt,linecolor=gray]{A0}{B3}
      \ncline[linewidth=0.5pt,linecolor=gray]{A0}{B4}

      \ncline[linewidth=0.5pt,linecolor=gray]{A0}{C1}
      \ncline[linewidth=0.5pt,linecolor=gray]{A0}{C2}
      \ncline[linewidth=0.5pt,linecolor=gray]{A0}{C3}
      \ncline[linewidth=0.5pt,linecolor=gray]{A0}{C4}

      \ncline[linewidth=0.5pt,linecolor=gray]{A0}{D1}
      \ncline[linewidth=0.5pt,linecolor=gray]{A0}{D2}
      \ncline[linewidth=0.5pt,linecolor=gray]{A0}{D3}
      \ncline[linewidth=0.5pt,linecolor=gray]{A0}{D4}

      \ncline[linewidth=0.5pt,linecolor=gray]{A0}{E1}
      \ncline[linewidth=0.5pt,linecolor=gray]{A0}{E2}
      \ncline[linewidth=0.5pt,linecolor=gray]{A0}{E3}
      \ncline[linewidth=0.5pt,linecolor=gray]{A0}{E4}

      \ncline[linewidth=0.5pt,linecolor=gray]{A0}{A1}
      \ncline[linewidth=0.5pt,linecolor=gray]{A0}{A2}
      \ncline[linewidth=0.5pt,linecolor=gray]{A0}{A3}
      \ncline[linewidth=0.5pt,linecolor=gray]{A0}{A4}

      \ncline[linewidth=0.5pt,linecolor=gray]{B0}{B1}
      \ncline[linewidth=0.5pt,linecolor=gray]{B0}{B2}
      \ncline[linewidth=0.5pt,linecolor=gray]{B0}{B3}
      \ncline[linewidth=0.5pt,linecolor=gray]{B0}{B4}

      \ncline[linewidth=0.5pt,linecolor=gray]{C0}{C1}
      \ncline[linewidth=0.5pt,linecolor=gray]{C0}{C2}
      \ncline[linewidth=0.5pt,linecolor=gray]{C0}{C3}
      \ncline[linewidth=0.5pt,linecolor=gray]{C0}{C4}

      \ncline[linewidth=0.5pt,linecolor=gray]{D0}{D1}
      \ncline[linewidth=0.5pt,linecolor=gray]{D0}{D2}
      \ncline[linewidth=0.5pt,linecolor=gray]{D0}{D3}
      \ncline[linewidth=0.5pt,linecolor=gray]{D0}{D4}

      \ncline[linewidth=0.5pt,linecolor=gray]{E0}{E1}
      \ncline[linewidth=0.5pt,linecolor=gray]{E0}{E2}
      \ncline[linewidth=0.5pt,linecolor=gray]{E0}{E3}
      \ncline[linewidth=0.5pt,linecolor=gray]{E0}{E4}

      \ncline[linewidth=0.5pt,linecolor=gray]{B1}{B2}
      \ncline[linewidth=0.5pt,linecolor=gray]{B1}{B3}
      \ncline[linewidth=0.5pt,linecolor=gray]{B1}{B4}
      \ncline[linewidth=0.5pt,linecolor=gray]{B2}{B3}
      \ncline[linewidth=0.5pt,linecolor=gray]{B2}{B4}
      \ncline[linewidth=0.5pt,linecolor=gray]{B3}{B4}

      \ncline[linewidth=0.5pt,linecolor=gray]{C1}{C2}
      \ncline[linewidth=0.5pt,linecolor=gray]{C1}{C3}
      \ncline[linewidth=0.5pt,linecolor=gray]{C1}{C4}
      \ncline[linewidth=0.5pt,linecolor=gray]{C2}{C3}
      \ncline[linewidth=0.5pt,linecolor=gray]{C2}{C4}
      \ncline[linewidth=0.5pt,linecolor=gray]{C3}{C4}

      \ncline[linewidth=0.5pt,linecolor=gray]{D1}{D2}
      \ncline[linewidth=0.5pt,linecolor=gray]{D1}{D3}
      \ncline[linewidth=0.5pt,linecolor=gray]{D1}{D4}
      \ncline[linewidth=0.5pt,linecolor=gray]{D2}{D3}
      \ncline[linewidth=0.5pt,linecolor=gray]{D2}{D4}
      \ncline[linewidth=0.5pt,linecolor=gray]{D3}{D4}

      \ncline[linewidth=0.5pt,linecolor=gray]{E1}{E2}
      \ncline[linewidth=0.5pt,linecolor=gray]{E1}{E3}
      \ncline[linewidth=0.5pt,linecolor=gray]{E1}{E4}
      \ncline[linewidth=0.5pt,linecolor=gray]{E2}{E3}
      \ncline[linewidth=0.5pt,linecolor=gray]{E2}{E4}
      \ncline[linewidth=0.5pt,linecolor=gray]{E3}{E4}

      \ncline[linewidth=0.5pt,linecolor=gray]{A1}{A2}
      \ncline[linewidth=0.5pt,linecolor=gray]{A1}{A3}
      \ncline[linewidth=0.5pt,linecolor=gray]{A1}{A4}
      \ncline[linewidth=0.5pt,linecolor=gray]{A2}{A3}
      \ncline[linewidth=0.5pt,linecolor=gray]{A2}{A4}
      \ncline[linewidth=0.5pt,linecolor=gray]{A3}{A4}

      \psdots( 0.00,-0.59)(-2.61, 1.81)( 2.61, 1.81)(-1.50, 2.76)( 1.50, 2.76)
      \psdots(-0.70,-1.79)( 0.70,-1.79)(-0.60,-1.34)( 0.60,-1.34)
      \psdots(-2.00, 0.00)(-4.00, 0.00)(-1.55, 0.90)(-3.10, 0.90)
      \psdots( 2.00, 0.00)( 4.00, 0.00)( 1.55, 0.90)( 3.10, 0.90)
      \psdots(-1.08, 1.85)(-2.14, 1.85)(-0.92, 2.14)(-1.86, 2.14)
      \psdots( 1.08, 1.85)( 2.14, 1.85)( 0.92, 2.14)( 1.86, 2.14)

      \put( 0.10,-0.64){\makebox(0,0)[l]{\color{color4} A0}}
      \put(-0.78,-1.79){\makebox(0,0)[r]{\color{color2} A1}}
      \put( 0.78,-1.79){\makebox(0,0)[l]{\color{color2} A2}}
      \put(-0.68,-1.34){\makebox(0,0)[r]{\color{color2} A3}}
      \put( 0.68,-1.34){\makebox(0,0)[l]{\color{color2} A4}}

      \put(-2.69, 1.81){\makebox(0,0)[r]{\color{color3} B0}}
      \put(-4.08, 0.00){\makebox(0,0)[r]{\color{color1} B1}}
      \put(-1.92, 0.05){\makebox(0,0)[l]{\color{color1} B2}}
      \put(-3.18, 0.90){\makebox(0,0)[r]{\color{color1} B3}}
      \put(-1.47, 0.90){\makebox(0,0)[l]{\color{color1} B4}}

      \put( 2.69, 1.81){\makebox(0,0)[l]{\color{color3} C0}}
      \put( 1.92, 0.05){\makebox(0,0)[r]{\color{color1} C1}}
      \put( 4.08, 0.00){\makebox(0,0)[l]{\color{color1} C2}}
      \put( 1.47, 0.90){\makebox(0,0)[r]{\color{color1} C3}}
      \put( 3.18, 0.90){\makebox(0,0)[l]{\color{color1} C4}}

      \put(-1.58, 2.76){\makebox(0,0)[r]{\color{color3} D0}}
      \put(-2.22, 1.85){\makebox(0,0)[r]{\color{color1} D1}}
      \put(-1.00, 1.85){\makebox(0,0)[l]{\color{color1} D2}}
      \put(-1.94, 2.14){\makebox(0,0)[r]{\color{color1} D3}}
      \put(-0.84, 2.14){\makebox(0,0)[l]{\color{color1} D4}}

      \put( 1.58, 2.76){\makebox(0,0)[l]{\color{color3} E0}}
      \put( 1.00, 1.85){\makebox(0,0)[r]{\color{color1} E1}}
      \put( 2.22, 1.85){\makebox(0,0)[l]{\color{color1} E2}}
      \put( 0.84, 2.14){\makebox(0,0)[r]{\color{color1} E3}}
      \put( 1.94, 2.14){\makebox(0,0)[l]{\color{color1} E4}}

      \end{pspicture}

    }
    \hfill\mbox{}
  \item[] (b)
  \item[]
    \mbox{}\hfill
    %\mbox{\includegraphics*[width=.8\textwidth]{fig3b.eps}}

      \begin{pspicture}(-6.0,-2.1)( 6.0, 2.1)
      \tiny

      %\psframe[linecolor=red](-6.0,-2.1)( 6.0, 2.1)

      \pnode(-5.0, 1.1){F25_top}
      \pnode(-5.0, 0.7){F25_bot}
      \pnode(-5.0, 0.5){F06_top}
      \pnode(-5.0, 0.1){F06_bot}
      \pnode(-5.0,-0.1){F05_top}
      \pnode(-5.0,-0.5){F05_bot}
      \pnode(-5.0,-0.7){F01_top}
      \pnode(-5.0,-1.1){F01_bot}

      \pnode(-5.5, 2.00){R25_max}
      \pnode(-5.5, 0.53){R25_min}
      \pnode(-5.5, 0.50){R06_max}
      \pnode(-5.5, 0.19){R06_min}
      \pnode(-5.5, 0.19){R05_max}
      \pnode(-5.5,-0.50){R05_min}
      \pnode(-5.5,-0.83){R01_max}
      \pnode(-5.5,-2.00){R01_min}

      \pspolygon[fillstyle=solid,fillcolor=lightgray,linestyle=none](-5.0, 1.1)(-5.5, 2.00)(-5.5, 0.53)(-5.0, 0.7)(-5.0, 1.1)
      \pspolygon[fillstyle=solid,fillcolor=lightgray,linestyle=none](-5.0, 0.5)(-5.5, 0.50)(-5.5, 0.19)(-5.0, 0.1)(-5.0, 0.5)
      \pspolygon[fillstyle=solid,fillcolor=lightgray,linestyle=none](-5.0,-0.1)(-5.5, 0.19)(-5.5,-0.50)(-5.0,-0.5)(-5.0,-0.1)
      \pspolygon[fillstyle=solid,fillcolor=lightgray,linestyle=none](-5.0,-0.7)(-5.5,-0.83)(-5.5,-2.00)(-5.0,-1.1)(-5.0,-0.7)

      \ncline[linewidth=0.4pt,linestyle=dashed]{F25_top}{R25_max}
      \ncline[linewidth=0.4pt]{F25_bot}{R25_min}
      \ncline[linewidth=0.4pt]{F06_top}{R06_max}
      \ncline[linewidth=0.4pt]{F06_bot}{R06_min}
      \ncline[linewidth=0.4pt]{F05_top}{R05_max}
      \ncline[linewidth=0.4pt]{F05_bot}{R05_min}
      \ncline[linewidth=0.4pt]{F01_top}{R01_max}
      \ncline[linewidth=0.4pt,linestyle=dashed]{F01_bot}{R01_min}

      \rput{L}(-5.7, 0.0){\scriptsize resistance}
      \psline[linewidth=0.8pt,arrowsize=1.5pt 4]{->}(-5.5,-2.0)(-5.5, 2.0)

      \pnode( 5.0, 1.1){G25_top}
      \pnode( 5.0, 0.7){G25_bot}
      \pnode( 5.0, 0.5){G06_top}
      \pnode( 5.0, 0.1){G06_bot}
      \pnode( 5.0,-0.1){G05_top}
      \pnode( 5.0,-0.5){G05_bot}
      \pnode( 5.0,-0.7){G01_top}
      \pnode( 5.0,-1.1){G01_bot}

      \pnode( 5.5, 2.00){T25_min}
      \pnode( 5.5, 0.66){T25_max}
      \pnode( 5.5, 0.50){T06_min}
      \pnode( 5.5,-0.34){T06_max}
      \pnode( 5.5,-0.34){T05_min}
      \pnode( 5.5,-0.50){T05_max}
      \pnode( 5.5,-0.50){T01_min}
      \pnode( 5.5,-2.00){T01_max}

      \pspolygon[fillstyle=solid,fillcolor=lightgray,linestyle=none]( 5.0, 1.1)( 5.5, 2.00)( 5.5, 0.66)( 5.0, 0.7)( 5.0, 1.1)
      \pspolygon[fillstyle=solid,fillcolor=lightgray,linestyle=none]( 5.0, 0.5)( 5.5, 0.50)( 5.5,-0.34)( 5.0, 0.1)( 5.0, 0.5)
      \pspolygon[fillstyle=solid,fillcolor=lightgray,linestyle=none]( 5.0,-0.1)( 5.5,-0.34)( 5.5,-0.50)( 5.0,-0.5)( 5.0,-0.1)
      \pspolygon[fillstyle=solid,fillcolor=lightgray,linestyle=none]( 5.0,-0.7)( 5.5,-0.50)( 5.5,-2.00)( 5.0,-1.1)( 5.0,-0.7)

      \ncline[linewidth=0.4pt,linestyle=dashed]{G25_top}{T25_min}
      \ncline[linewidth=0.4pt]{G25_bot}{T25_max}
      \ncline[linewidth=0.4pt]{G06_top}{T06_min}
      \ncline[linewidth=0.4pt]{G06_bot}{T06_max}
      \ncline[linewidth=0.4pt]{G05_top}{T05_min}
      \ncline[linewidth=0.4pt]{G05_bot}{T05_max}
      \ncline[linewidth=0.4pt]{G01_top}{T01_min}
      \ncline[linewidth=0.4pt,linestyle=dashed]{G01_bot}{T01_max}

      \rput{R}( 5.7, 0.0){\scriptsize time}
      \psline[linewidth=0.8pt,arrowsize=1.5pt 4]{->}( 5.5, 2.0)( 5.5,-2.0)

      \put(-4.8, 0.9){\makebox(0,0){\color{color4} A0}}
      \put(-4.4, 0.9){\makebox(0,0){\color{color2} A1}}
      \put(-4.0, 0.9){\makebox(0,0){\color{color2} A2}}
      \put(-3.6, 0.9){\makebox(0,0){\color{color2} A3}}
      \put(-3.2, 0.9){\makebox(0,0){\color{color2} A4}}
      \put(-2.8, 0.9){\makebox(0,0){\color{color3} B0}}
      \put(-2.4, 0.9){\makebox(0,0){\color{color1} B1}}
      \put(-2.0, 0.9){\makebox(0,0){\color{color1} B2}}
      \put(-1.6, 0.9){\makebox(0,0){\color{color1} B3}}
      \put(-1.2, 0.9){\makebox(0,0){\color{color1} B4}}
      \put(-0.8, 0.9){\makebox(0,0){\color{color3} C0}}
      \put(-0.4, 0.9){\makebox(0,0){\color{color1} C1}}
      \put( 0.0, 0.9){\makebox(0,0){\color{color1} C2}}
      \put( 0.4, 0.9){\makebox(0,0){\color{color1} C3}}
      \put( 0.8, 0.9){\makebox(0,0){\color{color1} C4}}
      \put( 1.2, 0.9){\makebox(0,0){\color{color3} D0}}
      \put( 1.6, 0.9){\makebox(0,0){\color{color1} D1}}
      \put( 2.0, 0.9){\makebox(0,0){\color{color1} D2}}
      \put( 2.4, 0.9){\makebox(0,0){\color{color1} D3}}
      \put( 2.8, 0.9){\makebox(0,0){\color{color1} D4}}
      \put( 3.2, 0.9){\makebox(0,0){\color{color3} E0}}
      \put( 3.6, 0.9){\makebox(0,0){\color{color1} E1}}
      \put( 4.0, 0.9){\makebox(0,0){\color{color1} E2}}
      \put( 4.4, 0.9){\makebox(0,0){\color{color1} E3}}
      \put( 4.8, 0.9){\makebox(0,0){\color{color1} E4}}
      \psline (-4.6, 0.7)(-4.6, 1.1)
      \psline (-4.2, 0.7)(-4.2, 1.1)
      \psline (-3.8, 0.7)(-3.8, 1.1)
      \psline (-3.4, 0.7)(-3.4, 1.1)
      \psline (-3.0, 0.7)(-3.0, 1.1)
      \psline (-2.6, 0.7)(-2.6, 1.1)
      \psline (-2.2, 0.7)(-2.2, 1.1)
      \psline (-1.8, 0.7)(-1.8, 1.1)
      \psline (-1.4, 0.7)(-1.4, 1.1)
      \psline (-1.0, 0.7)(-1.0, 1.1)
      \psline (-0.6, 0.7)(-0.6, 1.1)
      \psline (-0.2, 0.7)(-0.2, 1.1)
      \psline ( 0.2, 0.7)( 0.2, 1.1)
      \psline ( 0.6, 0.7)( 0.6, 1.1)
      \psline ( 1.0, 0.7)( 1.0, 1.1)
      \psline ( 1.4, 0.7)( 1.4, 1.1)
      \psline ( 1.8, 0.7)( 1.8, 1.1)
      \psline ( 2.2, 0.7)( 2.2, 1.1)
      \psline ( 2.6, 0.7)( 2.6, 1.1)
      \psline ( 3.0, 0.7)( 3.0, 1.1)
      \psline ( 3.4, 0.7)( 3.4, 1.1)
      \psline ( 3.8, 0.7)( 3.8, 1.1)
      \psline ( 4.2, 0.7)( 4.2, 1.1)
      \psline ( 4.6, 0.7)( 4.6, 1.1)
      \psframe(-5.0, 0.7)( 5.0, 1.1)

      \put(-4.8, 0.3){\makebox(0,0){\color{color4} A0}}
      \put(-4.4, 0.3){\makebox(0,0){\color{color2} A1}}
      \put(-4.0, 0.3){\makebox(0,0){\color{color2} A2}}
      \put(-3.6, 0.3){\makebox(0,0){\color{color2} A3}}
      \put(-3.2, 0.3){\makebox(0,0){\color{color2} A4}}
      \put(-2.8, 0.3){\makebox(0,0){\color{color3} B0}}
      \put(-2.4, 0.3){\makebox(0,0){\color{color1} B1}}
      \put(-2.0, 0.3){\makebox(0,0){\color{color1} B2}}
      \put(-1.6, 0.3){\makebox(0,0){\color{color1} B3}}
      \put(-1.2, 0.3){\makebox(0,0){\color{color1} B4}}
      \put(-0.8, 0.3){\makebox(0,0){\color{color3} C0}}
      \put(-0.4, 0.3){\makebox(0,0){\color{color1} C1}}
      \put( 0.0, 0.3){\makebox(0,0){\color{color1} C2}}
      \put( 0.4, 0.3){\makebox(0,0){\color{color1} C3}}
      \put( 0.8, 0.3){\makebox(0,0){\color{color1} C4}}
      \put( 1.2, 0.3){\makebox(0,0){\color{color3} D0}}
      \put( 1.6, 0.3){\makebox(0,0){\color{color1} D1}}
      \put( 2.0, 0.3){\makebox(0,0){\color{color1} D2}}
      \put( 2.4, 0.3){\makebox(0,0){\color{color1} D3}}
      \put( 2.8, 0.3){\makebox(0,0){\color{color1} D4}}
      \put( 3.2, 0.3){\makebox(0,0){\color{color3} E0}}
      \put( 3.6, 0.3){\makebox(0,0){\color{color1} E1}}
      \put( 4.0, 0.3){\makebox(0,0){\color{color1} E2}}
      \put( 4.4, 0.3){\makebox(0,0){\color{color1} E3}}
      \put( 4.8, 0.3){\makebox(0,0){\color{color1} E4}}
      \psline (-4.6, 0.1)(-4.6, 0.5)
      \psline (-3.0, 0.1)(-3.0, 0.5)
      \psline (-1.0, 0.1)(-1.0, 0.5)
      \psline ( 1.0, 0.1)( 1.0, 0.5)
      \psline ( 3.0, 0.1)( 3.0, 0.5)
      \psframe(-5.0, 0.1)( 5.0, 0.5)

      \put(-4.8,-0.3){\makebox(0,0){\color{color4} A0}}
      \put(-4.4,-0.3){\makebox(0,0){\color{color2} A1}}
      \put(-4.0,-0.3){\makebox(0,0){\color{color2} A2}}
      \put(-3.6,-0.3){\makebox(0,0){\color{color2} A3}}
      \put(-3.2,-0.3){\makebox(0,0){\color{color2} A4}}
      \put(-2.8,-0.3){\makebox(0,0){\color{color3} B0}}
      \put(-2.4,-0.3){\makebox(0,0){\color{color1} B1}}
      \put(-2.0,-0.3){\makebox(0,0){\color{color1} B2}}
      \put(-1.6,-0.3){\makebox(0,0){\color{color1} B3}}
      \put(-1.2,-0.3){\makebox(0,0){\color{color1} B4}}
      \put(-0.8,-0.3){\makebox(0,0){\color{color3} C0}}
      \put(-0.4,-0.3){\makebox(0,0){\color{color1} C1}}
      \put( 0.0,-0.3){\makebox(0,0){\color{color1} C2}}
      \put( 0.4,-0.3){\makebox(0,0){\color{color1} C3}}
      \put( 0.8,-0.3){\makebox(0,0){\color{color1} C4}}
      \put( 1.2,-0.3){\makebox(0,0){\color{color3} D0}}
      \put( 1.6,-0.3){\makebox(0,0){\color{color1} D1}}
      \put( 2.0,-0.3){\makebox(0,0){\color{color1} D2}}
      \put( 2.4,-0.3){\makebox(0,0){\color{color1} D3}}
      \put( 2.8,-0.3){\makebox(0,0){\color{color1} D4}}
      \put( 3.2,-0.3){\makebox(0,0){\color{color3} E0}}
      \put( 3.6,-0.3){\makebox(0,0){\color{color1} E1}}
      \put( 4.0,-0.3){\makebox(0,0){\color{color1} E2}}
      \put( 4.4,-0.3){\makebox(0,0){\color{color1} E3}}
      \put( 4.8,-0.3){\makebox(0,0){\color{color1} E4}}
      \psline (-3.0,-0.1)(-3.0,-0.5)
      \psline (-1.0,-0.1)(-1.0,-0.5)
      \psline ( 1.0,-0.1)( 1.0,-0.5)
      \psline ( 3.0,-0.1)( 3.0,-0.5)
      \psframe(-5.0,-0.1)( 5.0,-0.5)

      \put(-4.8,-0.9){\makebox(0,0){\color{color4} A0}}
      \put(-4.4,-0.9){\makebox(0,0){\color{color2} A1}}
      \put(-4.0,-0.9){\makebox(0,0){\color{color2} A2}}
      \put(-3.6,-0.9){\makebox(0,0){\color{color2} A3}}
      \put(-3.2,-0.9){\makebox(0,0){\color{color2} A4}}
      \put(-2.8,-0.9){\makebox(0,0){\color{color3} B0}}
      \put(-2.4,-0.9){\makebox(0,0){\color{color1} B1}}
      \put(-2.0,-0.9){\makebox(0,0){\color{color1} B2}}
      \put(-1.6,-0.9){\makebox(0,0){\color{color1} B3}}
      \put(-1.2,-0.9){\makebox(0,0){\color{color1} B4}}
      \put(-0.8,-0.9){\makebox(0,0){\color{color3} C0}}
      \put(-0.4,-0.9){\makebox(0,0){\color{color1} C1}}
      \put( 0.0,-0.9){\makebox(0,0){\color{color1} C2}}
      \put( 0.4,-0.9){\makebox(0,0){\color{color1} C3}}
      \put( 0.8,-0.9){\makebox(0,0){\color{color1} C4}}
      \put( 1.2,-0.9){\makebox(0,0){\color{color3} D0}}
      \put( 1.6,-0.9){\makebox(0,0){\color{color1} D1}}
      \put( 2.0,-0.9){\makebox(0,0){\color{color1} D2}}
      \put( 2.4,-0.9){\makebox(0,0){\color{color1} D3}}
      \put( 2.8,-0.9){\makebox(0,0){\color{color1} D4}}
      \put( 3.2,-0.9){\makebox(0,0){\color{color3} E0}}
      \put( 3.6,-0.9){\makebox(0,0){\color{color1} E1}}
      \put( 4.0,-0.9){\makebox(0,0){\color{color1} E2}}
      \put( 4.4,-0.9){\makebox(0,0){\color{color1} E3}}
      \put( 4.8,-0.9){\makebox(0,0){\color{color1} E4}}
      \psframe(-5.0,-0.7)( 5.0,-1.1)

      \end{pspicture}

    \hfill\mbox{}
  \end{indented}
  \caption{The network structure (a) corresponds to the hierarchical network proposed by Ravasz and Barabasi (see text for details). In (b), communities found at different scales for the network depicted. The left arrow represents the value of the resistance for which these structures prevail. The right arrow stands for the time intervals for which the same structures are found in a synchronization process (see text for details). For large positive values of $r$ the network is decomposed in individual nodes, while for large negative values of $r$ the whole network forms a single community.}
  \label{fig3}
\end{figure}
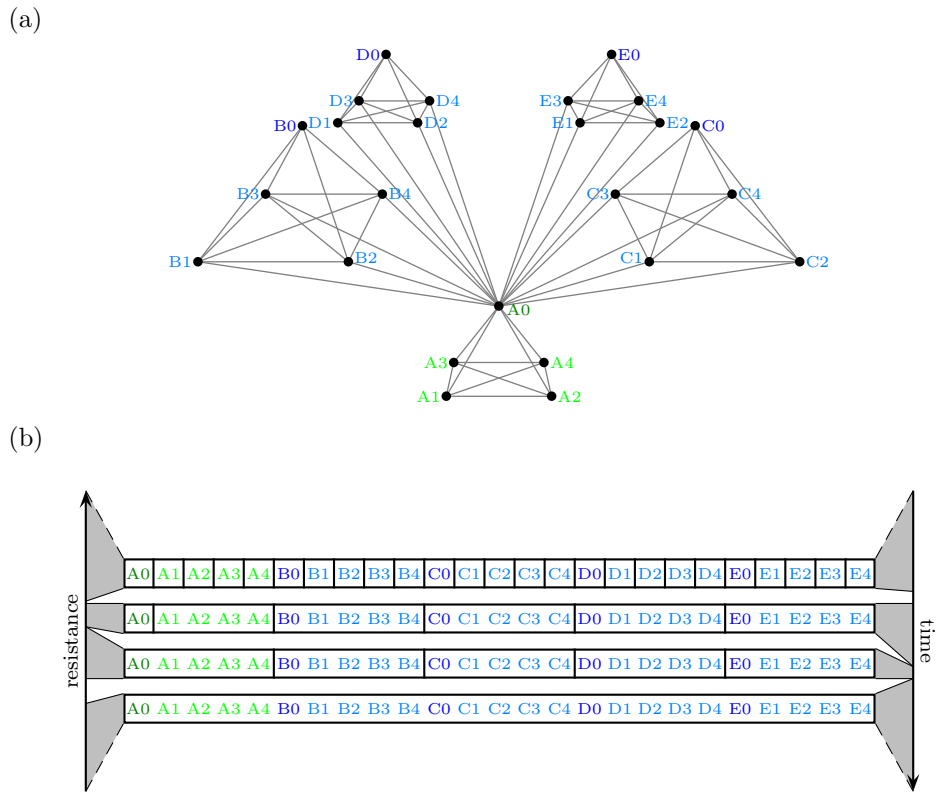

\begin{figure}[t]%figure4
  \begin{indented}
  \item[] (a)\hspace{6.2cm}(b)
  \item[]
    \mbox{}\hfill
    \mbox{\includegraphics*[width=.7\textwidth,height=.85\textwidth,angle=-90]{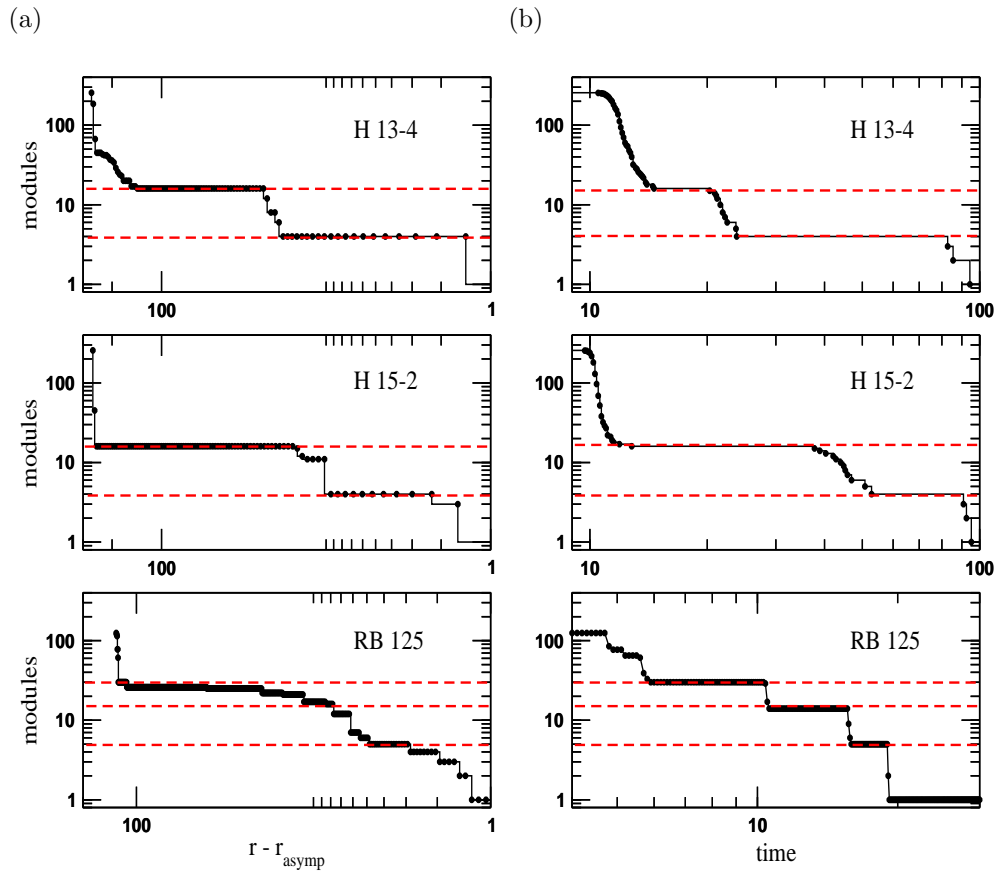}}
    \hfill\mbox{}
  \end{indented}
  \caption{Comparison between topological scales and dynamical scales of synchronization. The plots represent, in log-log scale, the number of communities as a function of (a) the translated resistance $r-r_{\mbox{\sz asymp}}$, and (b) time. Dashed red lines are a guide to the eye to emphasize the correspondence between the plateaus observed. Legends refer to the network structure (see text for details).}
  \label{fig4}
\end{figure}

We show the corroboration of these claims in a set of synthetic networks, where the modular structure at different scales is imposed by construction. In figure~\ref{fig3}, we sketch first the topology of a simple model of hierarchical network \cite{rb}, and the comparison between the specific communities found at different resolution levels, and the synchronization patterns observed in the path towards synchronization. The synthetic network of 25 nodes used combines the scale-free property with a high clustering coefficient, and can be iterated, following the scheme plotted in figure~\ref{fig3}a, to have many hierarchical levels. The results of the comparison reveal a strong equivalence between both processes, the static resolution method at different scales (different values of the resistance), and the groups of synchronized nodes in time. In figure~\ref{fig4} we extend this comparison for three more network structures: H~13-4 corresponding to the homogeneous in degree network described in section~\ref{sec:results}; equivalently the H~15-2 network \cite{arenas} corresponds to a homogeneous in degree network with two predefined hierarchical levels, being 256 the number of nodes, 15 the number of links of each node with the most internal community, 2 the number of links with the most external community, and 1 more link with any other node at random in the network; the RB~125 network has been used also in section~\ref{sec:results} and corresponds to the same scheme exposed in figure~\ref{fig3}a adding a new hierarchical level. The plots here represent, in log-log scale, the number of communities as a function of the translated resistance $r-r_{\mbox{\sz asymp}}$, and time. The correspondence between the patterns is highlighted, and again the correspondence is overwhelming.

Obviously, the functioning of real complex networks can rely on dynamical processes very different from the synchronization process exposed here, however it is still instructive to see how a simple nonlinear process reflects the mesoscale of complex networks, or from another view point, to see how the topology of the networks imposes dynamical (temporal) scales in their functioning.

\subsection{Comparison with other methods}
Some authors have proposed algorithms to extract the hierarchical organization of complex networks by modifying the objective function \cite{bornholdt}, or by searching local minima of the modularity landscape \cite{sales}. These approaches differ conceptually from ours and also in practice: i) the modification of the quality function \cite{bornholdt} does not always provide with the correct substructure of networks; ii) the method based on the screening of local minima of modularity \cite{sales} is designed assuming that the structure is hierarchical, which is not the case in many real networks.

The method proposed by Sales-Pardo et al.\ is specifically designed to unravel the hierarchical structure in networks. The comparison with our method is then not possible within our more general scope of any topology. For hierarchical networks, their method will find the hierarchy of scales, as ours also does, however, for non-hierarchical networks, their method can only produce nested communities, in contrast with ours. Conceptually, the difference can be summarized as follows: hierarchies imply multiple scales of description, but the implication does not hold in reverse.

The method proposed by Reichardt and Bornholdt (R\&B) \cite{bornholdt} was not designed to avoid the resolution limit of modularity but to offer a way to connect modularity with statistical physics. The main idea by the authors is to tune the null model (i.e.\ change the quality function) and then to obtain other partitions by maximizing the new quality functions. Indeed in \cite{kerstez} the authors interestingly showed that the R\&B method has the same resolution problems envisioned in \cite{fortunato} for modularity. The problem is that the R\&B method consisting into varying $\gamma$ (the prefactor that multiplies the null model) is not equivalent to tune the resolution of Newman's modularity. The authors in \cite{kerstez} recognize that if the size distribution of the communities is broad, like in collaboration networks or school friendship networks, there is no single proper value of $\gamma$ for the optimal resolution. The main difference with our method is that, no matter the size distribution of communities to be broad or not, the rescaling of the topology method that we present finds all the topological structure correctly because it is designed to this end.

\begin{figure}%figure5
  \begin{indented}
  \item[] (a)
  \item[]
    \mbox{}\hfill
    \mbox{\includegraphics*[width=.6\textwidth]{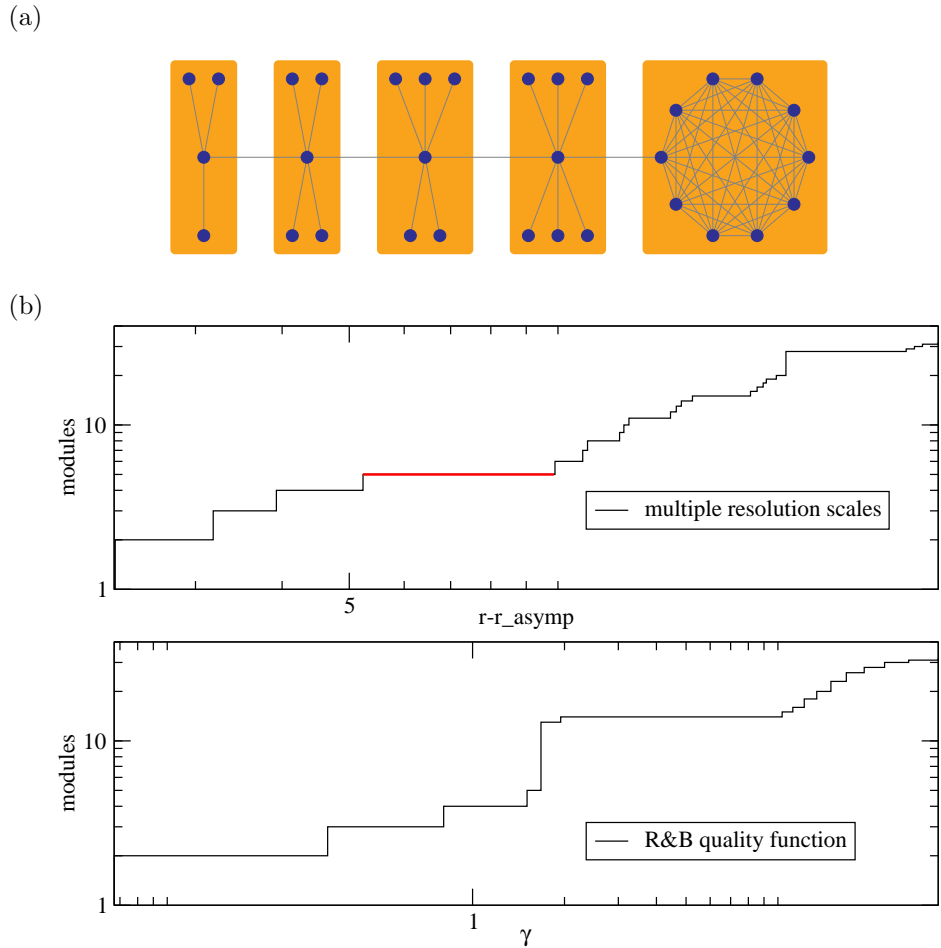}}
    \hfill\mbox{}
  \item[] (b)
  \item[]
    \mbox{}\hfill
    \mbox{\includegraphics*[width=.75\textwidth]{fig5b.eps}}
    \hfill\mbox{}
  \end{indented}
  \caption{(a) Toy model network with communities of different sizes and densities. (b) Top: number of modules as a function of the topological scale $r$ using our method; the red line indicates the range of scales for which the natural partition (four stars and clique) is obtained. Bottom: number of modules as a function of the parameter $\gamma$ in R\&B model; the natural partition (four stars and one clique) is not obtained for any value of $\gamma$.}
  \label{fig5}
\end{figure}

To support the above discussion we have built a toy model network with a simple topology but difficult for community detection algorithms because it includes communities of different sizes, some of them sparse and other dense, see figure~\ref{fig5}. The network model is small enough to have a clear vision of the modules, and to be attacked with computationally costly techniques in reasonable time. While our method succeeds in the process, the R\&B method fails. The results of our method and the R\&B method varying $\gamma$ are presented in figure~\ref{fig5}.

It is worth noticing that the parameter $\gamma$ in R\&B approach does not correspond to any value of $r$. Only when $\gamma=1$ and $r=0$ both definitions become equal, and are exactly Newman's original definition. Rewriting $Q_r$ in \req{QWSR} in terms of nodes,
\beq
  Q_r = \frac{1}{2w+Nr}\sum_i\sum_j \left(
      w_{ij} + r \delta_{ij} - \frac{(w_i+r)(w_j+r)}{2w+Nr}
    \right) \delta(C_i,C_j)\,,
\eeq
and comparing it to the R\&B modularity
\beq
  Q^{R\&B}_{\gamma} = \frac{1}{2w}\sum_i\sum_j \left(
      w_{ij} - \gamma \frac{w_i w_j}{2w}
    \right) \delta(C_i,C_j)\,,
\eeq
for both prescriptions to be equivalent for all partitions, one must show that
\beq
  \frac{(w_i+r)(w_j+r)}{2w+Nr} - r \delta_{ij} = \gamma \frac{w_i w_j}{2w}\,,\ \
  \forall i,j\,.
\eeq
If $i \neq j$, the relationship between $r$ and $\gamma$ becomes
\beq
  \gamma = \frac{2w}{2w+Nr}\, \cdot \frac{w_i+r}{w_i}\, \cdot \frac{w_j+r}{w_j}\,,
\eeq
which is only fulfilled for $r=0$ and $\gamma=1$, the trivial case stated before, because otherwise one would have different values of $\gamma$ for each pair of nodes $i$ and $j$. Therefore, there is no transformation of $Q_r$ into $Q^{R\&B}_{\gamma}$, and they are screening different things.

\subsection{Which is the ``best'' scale of description of complex networks?}
The question about the determination of the ``best'' scale of description of a complex network is natural, but ill posed in the current scenario. Throughout the paper we have stated that the more ``stable'' partitions, in terms of persistence maximizing $Q_r$ when varying the scales with $r$, are somehow more relevant in the topological description of the mesoscale. Their existence is an observed fact: some partitions are more persistent than others when changing the resolution scale of the topology. We think that this fact is not surprising, as it is not in many physical systems: phenomena that are observed persistently over a wide range of scales vanish at other scales, and others emerge. In general, these more persistent phenomena are usually more important to understand the system. More stable partitions are relevant in the sense that they usually have known meaning, but we cannot state that other partitions not so prevalent are uninformative. All them are embedded in the topology and give their particular information.

Summarizing what we think about the determination of the ``best''  or ``more relevant'' scale of description, we can say that the existence of relevant scales of description of a complex network should unavoidably pass through the definition of ``relevant''. Throughout the paper we have never tried to define ``relevant'' directly from the results obtained with our method, but {\em a posteriori}. We use, in the case of synthetic networks and real networks, the information that we have a priori (e.g.\ knowledge about the hierarchy imposed by construction, or known splits) to determine which scale is more relevant and then to check whether it is found by the method. What we observe is that these relevant scales are usually related to partitions that are significantly persistent (stable) at different scales (variation of $r$). However, to invert the argument is not straightforward. It is true that one could invent a function that peaks at the scales we see in reality that are known, as for example a function that accounts for the homogeneity of the obtained communities, but this inevitably imposes new conditions to the definition of module. Matching the discussion above let us expose the following: if it exists such a function that indicates the most ``relevant'' scale of description (and then partition), why not use this function as the objective function to optimize? This argument is strong because it implies that to determine if any scale is more important than others one must optimize a different quality function designed to this end, not modularity.

\section{Conclusions}
In conclusion, motivated by the recent finding that the optimization of modularity has a resolution limit, related to the characteristic scale imposed by the total strength (sum of weights) of the network, we propose a multiple resolution procedure that allows the optimization of modularity process to go deep into the structure. The main idea consists in to re-scale the topology by defining a new network from the original one, providing each node with a self-loop of the same magnitude $r$. The new network presents the same characteristics that the original network in terms of connectivity, but allows the search of modules at different topological scales. We have provided examples of the modular substructure found in synthetic and real complex networks. The results are sets of partitions that screen the full range of structural modules from individual nodes up to the whole network in each particular scale.

The analysis of the results reveal that some topological scales are more persistent (stable) in terms of resolution than others. These stable scales provide with specific information about the main modular aspects of the structure: in the synthetic networks analyzed, they correspond to the predefined structural scales imposed {\em ad hoc}; and in real networks they correspond exactly to previous knowledge about the networks, that has not been recovered by any other method studying these network topologies up to now. With this method, we release optimization of modularity from resolution problems, and give new ideas about the description of complex networks. The existence of several scales of description of complex networks, has deep analogies with the common study of complex systems in physics, where different models have been formulated at different spatial scales to get insight in different aspects of their phenomenology.

\section*{Acknowledgements}
This work was supported by Spanish Ministry of Science and Technology Grant FIS2006-13321-C02-02. We acknowledge for the usage of the resources, technical expertise and assistance provided by BSC-CNS supercomputing facility.

\appendix

\section{Resistance limiting cases}
Here we present the mathematical proofs of the physical limiting cases of the resistance for weighted undirected and directed networks, namely the limit of resistance for which all nodes are isolated, and the limit for which all nodes become members of a single group that represents the whole network.

\subsection{Resistance limiting cases for weighted undirected networks}
Let $w_{ij}=w_{ji}\ge 0$, $i\neq j$, be the weights of a complex network, where $w_{ij}=0$ if there is no link between nodes $i$ and $j$. We suppose that this network is connected; otherwise, each connected component should be analyzed one by one. The addition of a common resistance $r$ to all nodes may be understood as the definition of a new network with weights
\beq
  w'_{ij} = \left\{
    \ba{lll}
      w_{ij} & \mbox{if} & i\neq j\,, \\
      r & \mbox{if} & i=j\,. \\
    \ea
  \right.
\eeq
The strengths of this network are
\beq
  w'_i = \sum_j w'_{ij} = w_i + r\,,
\eeq
and its total strength is
\beq
  2w'=\sum_i w'_i = 2w+Nr\,.
\eeq

Now, the modularity of the new network is calculated as
\beq
  Q_r = \frac{1}{2w'}\sum_{i}\sum_{j}\left(
           w'_{ij}-\frac{w'_i w'_j}{2w'}
           \right)\delta(C_i,C_j)\,,
  \label{QrNoDr}
\eeq
which may also be written as
\beq
  Q_r = \frac{1}{2w'}\sum_{i}\sum_{j(\neq i)}\left(
           w_{ij}-\frac{w'_i w'_j}{2w'}
           \right)\delta(C_i,C_j) + D_r\,,
  \label{QrDr}
\eeq
where
\beq
  D_r = \frac{1}{2w'}\sum_{i}\left(
         r-\frac{{w'_i}^2}{2w'}
         \right)\,.
\eeq
Note that $D_r$ does not depend on the community partition.

\subsubsection{All nodes isolated}
If we have
\beq
  w'_{ij}-\frac{w'_i w'_j}{2w'} < 0\,, \ \forall i \neq j\,,
\eeq
then $Q_r$ in \req{QrDr} is maximized when $\delta(C_i,C_j)=0\,, \forall i \neq j$, i.e.\ modularity attains its maximum when all nodes are isolated in clusters of just one node. In terms of the resistance they simply become second order inequalities,
\beq
  (2w+Nr) w_{ij} < (w_i+r) (w_j+r) \,, \ \forall i \neq j\,,
\eeq
which can easily be solved for all pairs of nodes joined by an edge. Thus, $r_{\mbox{\sz max}}$ is the minimum value of $r$ which satisfies all these inequalities, and for $r>r_{\mbox{\sz max}}$ all nodes are separated in the optimal community configuration.

\subsubsection{All nodes in the same community \label{allinone}}
Let us analyze the behaviour of modularity just to the right of the asymptote $r_{\mbox{\sz asymp}}=-\frac{2w}{N}$. For convenience, we write the resistance as
\beq
  r = -\frac{2w}{N} + \epsilon\,,
\eeq
where $\epsilon$ is a small positive constant.

The first term of modularity in \req{QrNoDr} can be split in the following way:
\bea
  \sum_i\sum_j\frac{w'_{ij}}{\epsilon}\delta(C_i,C_j) & = &
    \sum_i\sum_j\frac{w'_{ij}}{\epsilon} -
    \sum_i\sum_{j(\neq i)}\frac{w_{ij}}{\epsilon}(1-\delta(C_i,C_j)) \nn
  & = & 1-\frac{a}{\epsilon}\,,
\eea
being $a$ the sum of weights of edges connecting different communities. If there are two or more communities then $a>0$, otherwise $a=0$.

The analysis of the second (null case) term of \req{QrNoDr} requires a communities expansion:
\bea
  \sum_i\sum_j\frac{w'_i w'_j}{\epsilon^2}\delta(C_i,C_j) & = &
    \sum_c\left(
      \sum_i\sum_j\frac{w'_i w'_j}{\epsilon^2}\delta(C_i,c)\delta(C_j,c)
    \right) \nn
  & = & \sum_c\frac{1}{\epsilon^2}
    \left(\sum_i w'_i \delta(C_i,c)\right)
    \left(\sum_j w'_j \delta(C_j,c)\right) \nn
  & = & \frac{1}{\epsilon^2}\sum_c
    \left(\sum_i (w_i + r) \delta(C_i,c)\right)^2 \nn
  & = & \frac{b}{\epsilon^2}\,,
\eea
where $b>0$, and $b \sim O(\epsilon^2)$ only if all strengths are equal; on the contrary, $b\sim O(1)$.

Therefore,
\beq
  Q_{r_{\mbox{\tiny asymp}}+\epsilon} = 1 - \frac{a}{\epsilon}
                                          - \frac{b}{\epsilon^2}\,,
  \label{Qassymp}
\eeq
which has an asymptotic behavior
\beq
  \lim_{\epsilon \rightarrow 0^+} Q_{r_{\mbox{\tiny asymp}}+\epsilon} = \left\{
    \ba{ll}
      -\infty & \mbox{if two or more communities}\,, \\
      0       & \mbox{if only one community}\,. \\
    \ea
  \right.
\eeq
This means that, for values of the resistance just above the asymptote, the optimal communities configuration is that with all nodes together in a single module that corresponds to the whole network.

\subsection{Resistance limiting cases for weighted directed networks}
Let $w_{ij}\ge 0$, $i\neq j$ be the weight of an arc that goes from the $i$-th to the $j$-th node, where $w_{ij}=0$ if there is no link between them. We suppose that this network is connected in the weak sense (weak connected components), i.e.\ the connected components are found as if the arcs were undirected; otherwise, each connected component should be analyzed one by one.

The natural generalization of modularity to cope with directed networks was introduced in \cite{alexnjp}, and is expressed as
\beq
  Q_r = \frac{1}{2w}\sum_{i}\sum_{j}\left(
           w_{ij}-\frac{{w_i}^{\mbox{\sz out}}{w_j}^{\mbox{\sz in}}}{2w}
           \right)\delta(C_i,C_j)\,,
  \label{QrDir}
\eeq
where the input and output strengths of the network are
\bea
  w_i^{\mbox{\sz out}} &=& \sum_j w_{ij} \\
  w_j^{\mbox{\sz in}} &=& \sum_i w_{ij}
\eea
and its total strength is
\beq
  2w = \sum_i w_i^{\mbox{\sz out}} = \sum_j w_j^{\mbox{\sz in}}
     = \sum_{ij} w_{ij}\,.
\eeq

The addition of a common resistance $r$ to all nodes may be understood as the definition of a new network with weights
\beq
  w'_{ij} = \left\{
    \ba{lll}
      w_{ij} & \mbox{if} & i\neq j\,, \\
      r & \mbox{if} & i=j\,. \\
    \ea
  \right.
\eeq
The strengths of this network are
\bea
  {w'_i}^{\mbox{\sz out}} &=& w_i^{\mbox{\sz out}} + r\,,\\
  {w'_j}^{\mbox{\sz in}} &=& w_j^{\mbox{\sz in}} + r\,,
\eea
and its total strength is
\beq
  2w' = \sum_i {w'_i}^{\mbox{\sz out}} =
        \sum_j {w'_j}^{\mbox{\sz in}} = 2w+Nr\,.
\eeq

Now, the modularity \req{QrDir} of the new network is calculated as
\beq
  Q_r = \frac{1}{2w'}\sum_{i}\sum_{j}\left(
           w'_{ij}-\frac{{w'_i}^{\mbox{\sz out}}{w'_j}^{\mbox{\sz in}}}{2w'}
           \right)\delta(C_i,C_j)\,,
  \label{QrDirNoDr}
\eeq
which may also be written as
\beq
  Q_r = \frac{1}{2w'}\sum_{i}\sum_{j(\neq i)}\left(
           w_{ij}-\frac{{w'_i}^{\mbox{\sz out}}{w'_j}^{\mbox{\sz in}}}{2w'}
           \right)\delta(C_i,C_j) + D_r\,,
  \label{QrDirDr}
\eeq
where
\beq
  D_r = \frac{1}{2w'}\sum_{i}\left(
          r-\frac{{w'_i}^{\mbox{\sz out}}{w'_i}^{\mbox{\sz in}}}{2w'}
        \right)\,.
\eeq
Note that $D_r$ does not depend on the community partition.

\subsubsection{All nodes isolated}
If we have
\beq
  \left(
    w'_{ij}-\frac{{w'_i}^{\mbox{\sz out}}{w'_j}^{\mbox{\sz in}}}{2w'}
  \right) + \left(
    w'_{ji}-\frac{{w'_j}^{\mbox{\sz out}}{w'_i}^{\mbox{\sz in}}}{2w'}
  \right) < 0\,, \ \forall i < j\,,
\eeq
then $Q_r$ in \req{QrDirDr} is maximized when $\delta(C_i,C_j)=0\,, \forall i \neq j$, i.e.\ modularity attains its maximum when all nodes are isolated in clusters of just one node. In terms of the resistance they simply become second order inequalities,
\beq
  (2w+Nr) (w_{ij}+w_{ji}) <
    (w_i^{\mbox{\sz out}}+r) (w_j^{\mbox{\sz in}}+r) +
    (w_j^{\mbox{\sz out}}+r) (w_i^{\mbox{\sz in}}+r)\,, \ \forall i < j\,,
\eeq
which can easily be solved for all pairs of nodes joined by an arc. Thus, $r_{\mbox{\sz max}}$ is the minimum value of $r$ which satisfies all these inequalities, and for $r>r_{\mbox{\sz max}}$ all nodes are separated in the optimal community configuration.

\subsubsection{All nodes in the same community}
The analysis of this case follows the same steps as in \ref{allinone}, yielding also to \req{Qassymp}:
\beq
  Q_{r_{\mbox{\tiny asymp}}+\epsilon} = 1 - \frac{a}{\epsilon}
                                        - \frac{b}{\epsilon^2}\,.
\eeq
The only difference is that now:
\beq
  b = \sum_c
      \left(\sum_i (w_i^{\mbox{\sz out}}+r) \delta(C_i,c)\right)
      \left(\sum_j (w_j^{\mbox{\sz in}}+r) \delta(C_j,c)\right).
\eeq
Unlike for undirected networks, the value of $b$ is not guaranteed to be positive, and then:
\beq
  \lim_{\epsilon \rightarrow 0^+} Q_{r_{\mbox{\tiny asymp}}+\epsilon} = \left\{
    \ba{ll}
      -\sign{(b)}\ \infty & \mbox{if two or more communities}\,, \\
      0                   & \mbox{if only one community}\,. \\
    \ea
  \right.
\eeq
This means that, only if $b$ is positive for all the different community partitions, for values of the resistance just above the asymptote, the optimal communities configuration is that with all nodes together in a single module that corresponds to the whole network. Otherwise, the modularity will raise to $+\infty$ for the maximum modularity configuration, and the single module structure might not be present for any value of the resistance.

\section{Optimization of the modularity using the Tabu heuristic}
We propose a new method to optimize the modularity based on Tabu search \cite{tabu}. The algorithm proceeds as follows: starting from an initial solution (a partition in groups of nodes of the network), \textit{S\_Init}, an iterative process that explores the search space begins, stepping from the solution of the current iteration, \textit{S\_Iter}, to one of its neighbors, \textit{S\_Neig}. The neighborhood is composed by the partitions that are obtained from the current solution by the application of a local operator called {\em move}. In our case, the {\em move} operator acts on a node at a time moving it from its current community to another selected at random, or creating a new one. Among the solutions in the neighborhood the best one is chosen to become the new current solution for the next iteration of the algorithm.

In order to escape from local optima, a list of tabu moves is used. This tabu list stores and forbids the most recently accepted moves and it is updated as the algorithm proceeds, so that a move just added to the list is removed from it after a certain number of iterations (\textit{Tabu\_Tenure}) have passed. However, tabu moves are allowed when they lead to an improved solution. Once a solution is accepted, the node moved to obtain this solution is inserted into the tabu list, in order to prevent the movement of the same node during the next \textit{Tabu\_Tenure} iterations, unless this move leads us to the best solution found until that moment. We used a logarithmic function on the number of nodes as the number of idle iterations needed to stop the search.

\begin{tabbing}
88\=88\=88\=88\= \kill
\textbf{function} Tabu\_Modularity\_Optimization(Net: Network; S\_Init: Solution) \\
\> \textbf{returns} (S\_Best: Solution) \textbf{is} \\
\textbf{const} \\
\> Tabu\_Tenure: Natural := 5; \\
\textbf{var} \\
\> Tabu\_Moves: Array\_Of\_Natural; \{counters of forbidden moves\} \\
\> Max\_Idle: Natural; \{maximum number of idle iterations\} \\
\> Num\_Idle: Natural; \{number of idle iterations\} \\
\> S\_Iter: Solution; \{solution of the current iteration\} \\
\> S\_Neig: Solution; \{solution in the neighborhood\} \\
\> Node\_Best: Natural; \{node with the best move\} \\
\textbf{begin} \\
\> \textbf{for} Node := 1 \textbf{to} Number\_Of\_Nodes(Net) \textbf{do} \{initialize the tabu moves\} \\
\>\> Tabu\_Moves[Node] := 0; \\
\> \textbf{end for}; \\
\> Max\_Idle := Maximum\_Of\_Nonimprovements(Number\_Of\_Nodes(Net)); \\
\> Num\_Idle := 0; \\
\> S\_Iter := S\_Init; \\
\> S\_Best := S\_Init; \\
\> \textbf{while} Num\_Idle $<$ Max\_Idle \textbf{do} \\
\>\> Explore\_Neighborhood(Net, S\_Iter, S\_Best, Tabu\_Moves, S\_Neig, Node\_Best); \\
\>\> \textbf{for} Node := 1 \textbf{to} Number\_Of\_Nodes(Net) \textbf{do} \{decrease the tabu moves\} \\
\>\>\> Tabu\_Moves[Node] := Maximum(0, Tabu\_Moves[Node]-1); \\
\>\> \textbf{end for}; \\
\>\> Tabu\_Moves[Node\_Best] := Tabu\_Tenure; \\
\>\> S\_Iter := S\_Neig; \\
\>\> \textbf{if} Modularity(S\_Neig) $>$ Modularity(S\_Best) \textbf{then} \\
\>\>\> S\_Best := S\_Neig; \\
\>\>\> Num\_Idle := 0; \\
\>\> \textbf{else} \\
\>\>\> Num\_Idle := Num\_Idle + 1; \\
\>\> \textbf{end if}; \\
\> \textbf{end while}; \\
\> \textbf{return} S\_Best; \\
\textbf{end} Tabu\_Modularity\_Optimization; \\
\\
\textbf{procedure} Explore\_Neighborhood(Net: \textbf{in} Network; S\_Iter, S\_Best: \textbf{in} Solution; \\
\> Tabu\_Moves: \textbf{in out} Array\_Of\_Natural; \\
\> S\_Neig: \textbf{out} Solution; Node\_Best: \textbf{out} Natural) \textbf{is} \\
\textbf{var} \\
\> S\_Move: Solution; \{solution from the move of a node\} \\
\textbf{begin} \\
\> Node\_Best := 0; \\
\> \textbf{for} Node := 1 \textbf{to} Number\_Of\_Nodes(Net) \textbf{do} \\
\>\> S\_Move := Solution\_From\_Move(Net, S\_Iter, Node); \\
\>\> \textbf{if} Modularity(S\_Move) $>$ Modularity(S\_Best) \textbf{then} \\
\>\>\> Tabu\_Moves[Node] := 0; \\
\>\> \textbf{end if}; \\
\>\> \textbf{if} Tabu\_Moves[Node] = 0 \textbf{and} \\
\>\>\>\> (Node\_Best = 0 \textbf{or else} Modularity(S\_Move) $>$ Modularity(S\_Neig)) \textbf{then} \\
\>\>\> Node\_Best := Node; \\
\>\>\> S\_Neig := S\_Move; \\
\>\> \textbf{end if}; \\
\> \textbf{end for}; \\
\textbf{end} Explore\_Neighborhood; \\
\end{tabbing}

The main advantage of this algorithm is that it is a mixture of divisive and agglomerative processes, avoiding the drawbacks of both single strategies. Moreover, the iterative process can start from any initial partition, which is adequate for the mesoscale determination, since the optimal partitions for nearby values of the resistance are frequently similar. In terms of computational cost, the tabu heuristic is equivalent to other stochastic optimization methods such as simulated annealing or genetic algorithms.

\end{document}